\documentclass[]{pasj01}

\newcommand{\eg}{e.g., }

\newcommand{\Msun}{M_{\odot}}

\newcommand{\RADEC}{${\rm R.A.=13^{h}08^m}$, ${\rm decl.=-22^\circ30'}$ (J2000.0)}
\newcommand{\GW}{GW170817}

\def\gsim{\mathrel{\rlap{\lower 4pt \hbox{\hskip 1pt $\sim$}}\raise 1pt
\hbox {$>$}}}
\def\lsim{\mathrel{\rlap{\lower 4pt \hbox{\hskip 1pt $\sim$}}\raise 1pt
\hbox {$<$}}}

\newcommand{\NameA}{J-GEM17btc}

\newcommand{\NameC}{J-GEM17bog}
\newcommand{\ye}{Y_{\rm e}}

\begin{document} 
\Received{}
\Accepted{}

\title{Subaru Hyper Suprime-Cam Survey for An Optical Counterpart of GW170817}

\author{Nozomu \textsc{Tominaga}\altaffilmark{1,2}}
\author{Masaomi \textsc{Tanaka}\altaffilmark{3}}
\author{Tomoki \textsc{Morokuma}\altaffilmark{4}}
\author{Yousuke \textsc{Utsumi}\altaffilmark{5}}
\author{Masaki S. \textsc{Yamaguchi}\altaffilmark{4}}
\author{Naoki \textsc{Yasuda}\altaffilmark{2}}
\author{Masayuki \textsc{Tanaka}\altaffilmark{3}}
\author{Michitoshi \textsc{Yoshida}\altaffilmark{6}}
\author{Takuya \textsc{Fujiyoshi}\altaffilmark{6}}
\author{Hisanori \textsc{Furusawa}\altaffilmark{3}}
\author{Koji S. \textsc{Kawabata}\altaffilmark{5}}
\author{Chien-Hsiu \textsc{Lee}\altaffilmark{6}}
\author{Kentaro \textsc{Motohara}\altaffilmark{4}}
\author{Ryou \textsc{Ohsawa}\altaffilmark{4}}
\author{Kouji \textsc{Ohta}\altaffilmark{7}}
\author{Tsuyoshi \textsc{Terai}\altaffilmark{6}}
\author{Fumio \textsc{Abe}\altaffilmark{8}}
\author{Wako \textsc{Aoki}\altaffilmark{3}}
\author{Yuichiro \textsc{Asakura}\altaffilmark{8,$\dagger$}}
\author{Sudhanshu \textsc{Barway}\altaffilmark{9}}
\author{Ian A. \textsc{Bond}\altaffilmark{10}}
\author{Kenta \textsc{Fujisawa}\altaffilmark{0}}
\author{Satoshi \textsc{Honda}\altaffilmark{11}}
\author{Kunihito \textsc{Ioka}\altaffilmark{12}}
\author{Youichi \textsc{Itoh}\altaffilmark{11}}
\author{Nobuyuki \textsc{Kawai}\altaffilmark{13}}
\author{Ji Hoon \textsc{Kim}\altaffilmark{6}}
\author{Naoki \textsc{Koshimoto}\altaffilmark{14}}
\author{Kazuya \textsc{Matsubayashi}\altaffilmark{15}}
\author{Shota \textsc{Miyazaki}\altaffilmark{14}}
\author{Tomoki \textsc{Saito}\altaffilmark{11}}
\author{Yuichiro \textsc{Sekiguchi}\altaffilmark{16,12}}
\author{Takahiro \textsc{Sumi}\altaffilmark{14}}
\author{Paul J. \textsc{Tristram}\altaffilmark{17}}
\altaffiltext{1}{Department of Physics, Faculty of Science and Engineering, Konan University, 8-9-1 Okamoto, Kobe, Hyogo 658-8501, Japan}
\altaffiltext{2}{Kavli Institute for the Physics and Mathematics of the Universe (WPI), The University of Tokyo Institutes for Advanced Study, The University of Tokyo, 5-1-5 Kashiwa, Chiba 277-8583, Japan}
\altaffiltext{3}{National Astronomical Observatory of Japan, 2-21-1 Osawa, Mitaka, Tokyo 181-8588, Japan}
\altaffiltext{4}{Institute of Astronomy, Graduate School of Science, The University of Tokyo, 2-21-1 Osawa, Mitaka, Tokyo 181-0015, Japan}
\altaffiltext{5}{Hiroshima Astrophysical Science Center, Hiroshima University, 1-3-1 Kagamiyama, Higashi-Hiroshima, Hiroshima, 739-8526, Japan}
\altaffiltext{6}{Subaru Telescope, National Astronomical Observatory of Japan, 650 North A'ohoku Place, Hilo, HI 96720, USA}
\altaffiltext{7}{Department of Astronomy, Kyoto University, Kitashirakawa-Oiwake-cho, Sakyo-ku, Kyoto,  606-8502, Japan}
\altaffiltext{8}{Institute for Space-Earth Environmental Research, Nagoya University, Furo-cho, Chikusa, Nagoya, Aichi 464-8601, Japan}
\altaffiltext{9}{South African Astronomical Observatory, PO Box 9, 7935 Observatory, Cape Town, South Africa}
\altaffiltext{10}{Institute for Natural and Mathematical Sciences, Massey University, Private Bag 102904 North Shore Mail Centre, Auckland 0745, New Zealand}
\altaffiltext{11}{Nishi-Harima Astronomical Observatory, Center for Astronomy, University of Hyogo, 407-2, Nishigaichi, Sayo, Hyogo 679-5313, Japan}
\altaffiltext{12}{Center for Gravitational Physics, Yukawa Institute for Theoretical Physics, Kyoto University, Kyoto 606-8502, Japan}
\altaffiltext{13}{Department of Physics, Tokyo Institute of Technology, 2-12-1 Ookayama, Meguro-ku, Tokyo 152-8551, Japan}
\altaffiltext{14}{Department of Earth and Space Science, Graduate School of Science, Osaka University, 1-1 Machikaneyama, Toyonake, Osaka 560-0043, Japan}
\altaffiltext{15}{Okayama Astrophysical Observatory, National Astronomical Observatory of Japan, 3037-5 Honjo, Kamogata, Asakuchi, Okayama 719-0232, Japan}
\altaffiltext{16}{Department of Physics, Toho University, Funabashi, Chiba 274-8510, Japan}
\altaffiltext{17}{University of Canterbury, Mt John Observatory, PO Box 56, Lake Tekapo 7945, New Zealand}
\author{the J-GEM collaboration}
\email{tominaga@konan-u.ac.jp}

\altaffiltext{$\dagger$}{Deceased 18 August 2017}



\maketitle

\begin{abstract}
 We perform a $z$-band survey for an optical counterpart of a binary neutron star
 coalescence \GW\ with
 Subaru/Hyper Suprime-Cam. Our untargeted transient search covers $23.6$~deg$^2$
 corresponding to the $56.6\%$ credible region of \GW\ and reaches
 the $50\%$ completeness magnitude of $20.6$~mag on average. As a result, we find 60
 candidates of extragalactic transients, including \NameA\ (a.k.a. SSS17a/DLT17ck). While \NameA\ is
 associated with NGC~4993 that is firmly located inside the 3D skymap of \GW, the
 other 59 candidates do not have distance information in the GLADE v2
 catalog or NASA/IPAC Extragalactic Database (NED). Among 59 candidates,
 58 are located at the center of extended objects in the Pan-STARRS1
 catalog, while one candidate has an offset. We present location,
 $z$-band apparent magnitude, and time
 variability of the candidates and evaluate
 the probabilities that they are located
 inside of the
 3D skymap of \GW. The probability for \NameA\ is $64\%$ being much higher
 than those for the other 59 candidates ($9.3\times10^{-3}-2.1\times10^{-1}\%$). Furthermore, the possibility, that at least one
 of the other 59 candidates is located within the 3D skymap, is only $3.2\%$. Therefore, we conclude
 that \NameA\ is the most-likely and distinguished
 candidate as the optical counterpart of \GW. 
\end{abstract}

\section{Introduction}

The existence of gravitational waves (GWs) is predicted in the theory of
general relativity.
Although the existence is indirectly demonstrated by
the energy loss of a binary pulsar system \citep{hul75,tay82}, the direct
observation of GWs had not been realized owing to its small amplitudes. 
The first direct detection is
achieved with the Advanced Laser Interferometer Gravitational-Wave Observatory
(LIGO) on Sep 14, 2015 \citep{gw150914}. The first GW source
originates from the coalescence of two black holes, each
$\sim30\Msun$. The discovery is important not only for the direct
probe of the strong field dynamics of general relativity, but also
for the first evidence of a black
hole binary. LIGO and Advanced Virgo subsequently detect three GW signals
and one candidate signal, all from the coalescence of
black-hole binaries \citep{gw151226,lvt151012,gw170104,gw170814}. These
discoveries open the era of 
``gravitational wave astronomy''.

However, the conclusive identification of the GW sources on the sky
remains challenging because of  the poor sky localization with the gravitational
wave observations. The sky localization areas of four GW sources are about
$230-1160$~deg$^2$ ($90\%$ credible region) with two detectors of LIGO and about
$60$~deg$^2$ ($90\%$ credible region) even with three detectors including the Advanced Virgo.
Since there are many galaxies in the area, it is impossible to
determine the host galaxy of a GW source only with the GW observations. Therefore, 
multi-wavelength searches for electromagnetic (EM) counterparts are
initiated after the alerts of GW detection from the LIGO-Virgo networks.
So far, no firm EM counterparts have been found (\eg \cite{DECamgw150914,PSgw150914,PTFgw150914,JGEMgw150914,JGEMgw151226}),
except for a report of the putative detection for GW150914 with Fermi/GBM
(\cite{con16GW150914}, but questioned by
\cite{sav16GW150914,gre16GW150914}).

The non-detection of EM counterparts is not surprising because the four GWs are originated
from mergers of black holes, although several theoretical studies try to
explain the putative Fermi/GBM emission (e.g., \cite{yam16}). On the other
hand, first-principle numerical simulations with general relativity demonstrate that
binary coalescence including at least one neutron star (NS) can eject
materials as dynamical ejecta (e.g., \cite{ros99,gor11,hot13,bau13}) and post-merger ejecta (e.g., \cite{des09,fer13,shi17}).
The ejecta dominantly consist of $r$-process elements (e.g.,
\cite{lat74,eic89,kor12,wanajo14}), and thus the decay of radioactive
isotopes produced by the $r$-process nucleosynthesis heats up and
brightens the ejecta. The EM-bright object is called ``kilonova''
or ``macronova'' \citep{li98,kul05,met10}, and regarded as a promising EM
counterpart of a GW \citep{met12,kas13,bar13,tanaka13,met14,tanaka14,kas15,met17}.
Also, the central engine of a short gamma-ray burst, which is believed
to originate from a binary neutron star coalescence, is a possible energy
source of EM counterparts through its jet and gamma/X-ray emission (\eg
\cite{kis16}).

On Aug 17, 2017, 12:41:04 UTC, Advanced LIGO and Advanced Virgo detected
a GW candidate from a binary NS coalescence, being coincident
with a gamma-ray detection with Fermi/GBM
\citep{GW170817first,GW170817detection}. The sky localization with the three
detectors is as narrow as $28$~deg$^2$ for a $90\%$
credible region centered at \RADEC\
\citep{GW170817LVCpaper}. And the localization is
overlapped with the error regions of gamma-ray detection with Fermi/GBM
and INTEGRAL \citep{GW170817GBM,GW170817INTEGRAL,GW170817INTEGRALpaper}.
The GW observation reveals the luminosity distance to the GW source,
named \GW, as
$40^{+8}_{-14}$~Mpc ($90\%$ probability) \citep{GW170817LVCpaper}.
Although \GW\ appeared at the position close to the Sun, the first
significant alert of a binary NS coalescence and the narrow sky
localization area initiate many EM follow-up observations \citep{GW170817MMApaper}.

Along with the EM follow-up observation campaign of \GW, the Japanese
collaboration for Gravitational wave ElectroMagnetic follow-up (J-GEM)
performed a survey with Hyper Suprime-Cam (HSC, \cite{miy12}), which is a
wide-field imager installed on the prime focus of the 8.2m Subaru
telescope. Its FoV of $1.77$~deg$^2$ is largest among
the currently existing 8-10 m telescopes, and thus it is the most
efficient instrument for the optical survey. In this paper, we summarize
the observation with Subaru/HSC and properties of discovered candidates.
Throughout the paper, we correct the Galactic reddening 
\citep{sch11}\footnote{http://irsa.ipac.caltech.edu/applications/DUST/},
and all the magnitudes are given as AB magnitudes.

\begin{table}
  \tbl{Subaru/HSC pointings.}{%
  \begin{tabular}{ccc}
      \hline
      Pointing & R.A. & decl. \\
      (ID) & (J2000) & (J2000)\\ 
      \hline
04 & $13^{h}07^{m}25^{s}$ & $-26^\circ36'51''$ \\
05 & $13^{h}10^{m}14^{s}$ & $-27^\circ17'02''$ \\
06 & $13^{h}13^{m}03^{s}$ & $-27^\circ57'27''$ \\
07 & $13^{h}15^{m}51^{s}$ & $-28^\circ38'07''$ \\
08 & $13^{h}18^{m}40^{s}$ & $-29^\circ19'02''$ \\
09 & $13^{h}21^{m}29^{s}$ & $-30^\circ00'15''$ \\
10 & $13^{h}04^{m}36^{s}$ & $-24^\circ37'42''$ \\
11 & $13^{h}07^{m}25^{s}$ & $-25^\circ17'12''$ \\
12 & $13^{h}10^{m}14^{s}$ & $-25^\circ56'55''$ \\
13 & $13^{h}13^{m}03^{s}$ & $-26^\circ36'51''$ \\
14 & $13^{h}01^{m}48^{s}$ & $-22^\circ40'26''$ \\
15 & $13^{h}15^{m}51^{s}$ & $-27^\circ17'02''$ \\
16 & $13^{h}18^{m}40^{s}$ & $-27^\circ57'27''$ \\
17 & $13^{h}04^{m}36^{s}$ & $-23^\circ19'20''$ \\
18 & $13^{h}07^{m}25^{s}$ & $-23^\circ58'25''$ \\
19 & $12^{h}58^{m}59^{s}$ & $-20^\circ44'47''$ \\
20 & $13^{h}10^{m}14^{s}$ & $-24^\circ37'43''$ \\
22 & $13^{h}13^{m}03^{s}$ & $-25^\circ17'12''$ \\
23 & $13^{h}15^{m}51^{s}$ & $-25^\circ56'55''$ \\
24 & $12^{h}56^{m}10^{s}$ & $-18^\circ50'37''$ \\
25 & $13^{h}04^{m}36^{s}$ & $-22^\circ01'43''$ \\
26 & $13^{h}07^{m}25^{s}$ & $-22^\circ40'26''$ \\
28 & $13^{h}10^{m}14^{s}$ & $-23^\circ19'20''$ \\
29 & $13^{h}01^{m}48^{s}$ & $-20^\circ06'35''$ \\
      \hline
    \end{tabular}}\label{tab:pointing}
\end{table}

\section{Observation and data analysis}

\begin{figure}
 \begin{center}
  \includegraphics[width=8cm]{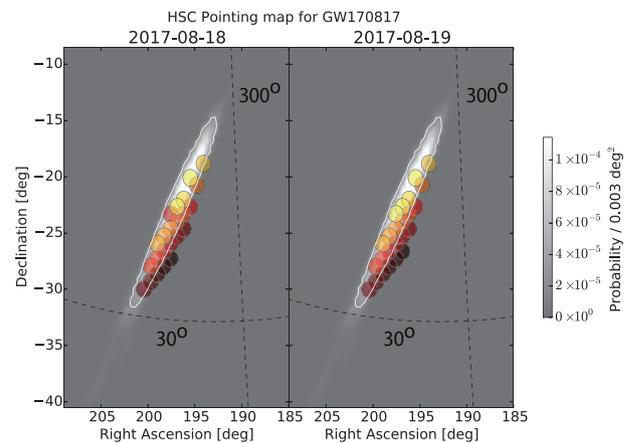} 
 \end{center}
\caption{Pointing map for GW170817 overlaid on the 
 probability map (LALInference\_v2.fits.gz;
 \cite{GW170817LVCpaper}). The white contour represents the $90\%$
 credible region. Circles represent the field-of-view of HSC,
 changing their face color with an order of observation. Observations
 have been carried out from darker color to lighter color.
 The dashed curves represent the Galactic graticules.}\label{fig:pointing}
\end{figure}

\begin{longtable}{ccc}
 \caption{Subaru/HSC observation log of exposures used in the analysis.}\label{tab:exposure}
 \hline
 Pointing & TaiObs & Exposure time\\
 & (UTC) & (s)\\ 
 \hline
 \endfirsthead
 \hline
 Pointing & TaiObs & Exposure time\\
 & (UTC) & (s)\\
 \hline
 \endhead
 \endfoot
 \endlastfoot
 28 & 2017-08-18T05:30:27 & 20.0 \\ 
 05 & 2017-08-18T05:32:00 & 30.0 \\ 
 06 & 2017-08-18T05:33:01 & 30.0 \\ 
 07 & 2017-08-18T05:34:03 & 30.0 \\ 
 08 & 2017-08-18T05:35:03 & 30.0 \\ 
 09 & 2017-08-18T05:36:03 & 30.0 \\ 
 10 & 2017-08-18T05:37:10 & 30.0 \\ 
 11 & 2017-08-18T05:38:10 & 30.0 \\ 
 12 & 2017-08-18T05:39:10 & 30.0 \\ 
 28 & 2017-08-18T05:40:11 & 60.0 \\ 
 13 & 2017-08-18T05:41:41 & 30.0 \\ 
 14 & 2017-08-18T05:42:42 & 30.0 \\ 
 15 & 2017-08-18T05:43:45 & 30.0 \\ 
 16 & 2017-08-18T05:44:46 & 30.0 \\ 
 17 & 2017-08-18T05:45:50 & 30.0 \\ 
 18 & 2017-08-18T05:46:51 & 30.0 \\ 
 19 & 2017-08-18T05:47:52 & 30.0 \\ 
 20 & 2017-08-18T05:48:54 & 30.0 \\ 
 22 & 2017-08-18T05:49:55 & 30.0 \\ 
 23 & 2017-08-18T05:50:55 & 30.0 \\ 
 24 & 2017-08-18T05:52:05 & 30.0 \\ 
 25 & 2017-08-18T05:53:09 & 30.0 \\ 
 26 & 2017-08-18T05:54:11 & 30.0 \\ 
 29 & 2017-08-18T05:55:16 & 30.0 \\ \hline 
 04 & 2017-08-19T05:22:21 & 10.0 \\ 
 05 & 2017-08-19T05:23:02 & 10.0 \\ 
 06 & 2017-08-19T05:23:46 & 10.0 \\ 
 07 & 2017-08-19T05:24:29 & 10.0 \\ 
 08 & 2017-08-19T05:25:11 & 10.0 \\ 
 09 & 2017-08-19T05:25:57 & 10.0 \\ 
 28 & 2017-08-19T05:26:44 & 30.0 \\ 
 10 & 2017-08-19T05:27:44 & 30.0 \\ 
 11 & 2017-08-19T05:28:45 & 30.0 \\ 
 12 & 2017-08-19T05:29:47 & 30.0 \\ 
 13 & 2017-08-19T05:30:49 & 30.0 \\ 
 28 & 2017-08-19T05:31:49 & 30.0 \\ 
 14 & 2017-08-19T05:32:51 & 30.0 \\ 
 15 & 2017-08-19T05:33:54 & 30.0 \\ 
 16 & 2017-08-19T05:34:55 & 30.0 \\ 
 17 & 2017-08-19T05:35:59 & 30.0 \\ 
 18 & 2017-08-19T05:37:01 & 30.0 \\ 
 19 & 2017-08-19T05:38:02 & 30.0 \\ 
 20 & 2017-08-19T05:39:03 & 30.0 \\ 
 22 & 2017-08-19T05:40:05 & 30.0 \\ 
 23 & 2017-08-19T05:41:06 & 30.0 \\ 
 24 & 2017-08-19T05:42:15 & 30.0 \\ 
 25 & 2017-08-19T05:43:15 & 30.0 \\ 
 26 & 2017-08-19T05:44:15 & 30.0 \\ 
 29 & 2017-08-19T05:45:16 & 30.0 \\ 
 28 & 2017-08-19T05:46:17 & 30.0 \\ \hline 
 28 & 2017-08-25T05:22:45 & 10.0 \\ 
 28 & 2017-08-25T05:23:26 & 10.0 \\ 
 28 & 2017-08-25T05:24:06 & 10.0 \\ 
 28 & 2017-08-25T05:24:48 & 10.0 \\ 
 28 & 2017-08-25T05:25:29 & 20.0 \\ 
 28 & 2017-08-25T05:27:10 & 20.0 \\ 
 28 & 2017-08-25T05:28:01 & 30.0 \\ 
 28 & 2017-08-25T05:29:07 & 30.0 \\ \hline 
 28 & 2017-08-27T05:24:07 & 10.0 \\ 
 28 & 2017-08-27T05:24:48 & 10.0 \\ 
 28 & 2017-08-27T05:25:28 & 10.0 \\ 
 28 & 2017-08-27T05:26:09 & 10.0 \\ 
 28 & 2017-08-27T05:26:49 & 10.0 \\ 
 28 & 2017-08-27T05:27:30 & 10.0 \\ 
 28 & 2017-08-27T05:28:10 & 10.0 \\ 
 28 & 2017-08-27T05:28:51 & 10.0 \\ 
 28 & 2017-08-27T05:29:32 & 10.0 \\ 
 28 & 2017-08-27T05:30:12 & 10.0 \\ 
 28 & 2017-08-27T05:30:53 & 10.0 \\ 
 28 & 2017-08-27T05:31:34 & 10.0 \\ 
 28 & 2017-08-27T05:32:15 & 20.0 \\
 \hline
\end{longtable}

\begin{figure}
 \begin{center}
  \includegraphics[width=8cm]{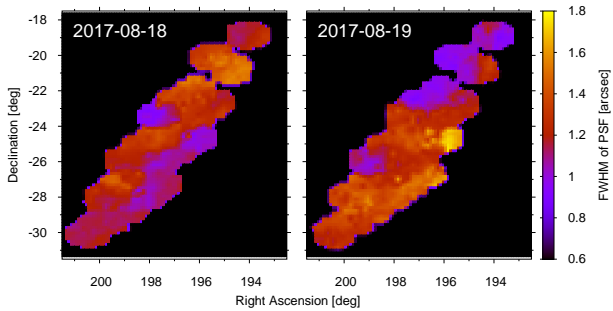} 
 \end{center}
\caption{Map of FWHM of PSF in the stacked images on Aug 18 and 19, 2017.}\label{fig:seeing}
\end{figure}

\begin{figure}
 \begin{center}
  \includegraphics[width=8cm]{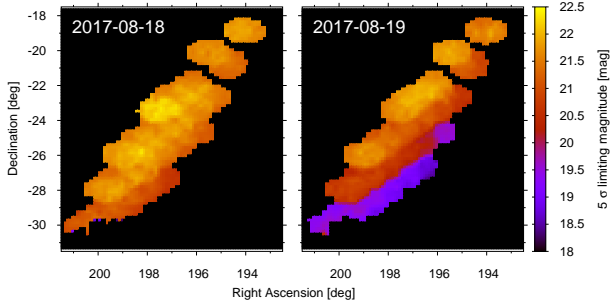} 
 \end{center}
\caption{Map of $5\sigma$ limiting magnitude in the difference images on Aug 18 and 19, 2017.}\label{fig:limmag}
\end{figure}

\begin{table}
  \tbl{Seeing of stacked images.}{%
  \begin{tabular}{cccc}
      \hline
      Date & \multicolumn{3}{c}{FWHM of PSF ('')}\\ 
      (UTC) & min & median & max\\ 
      \hline
      2017-08-18 & 0.91 & 1.20 & 1.62 \\
      2017-08-19 & 0.73 & 1.25 & 1.80 \\
      2017-08-25 & 0.75 & 0.90 & 1.16 \\
      2017-08-27 & 1.13 & 1.21 & 1.50 \\
      \hline
    \end{tabular}}\label{tab:seeing}
\end{table}

\begin{table}
  \tbl{$5\sigma$ limiting magnitude of difference images.}{%
  \begin{tabular}{cccc}
   \hline
   Date & \multicolumn{3}{c}{Limiting magnitude (mag)}\\ 
   (UTC) & min & median & max\\ 
   \hline
   2017-08-18 & 20.47 & 21.61 & 22.51 \\
   2017-08-19 & 18.30 & 20.97 & 22.21 \\
   2017-08-25 & 21.06 & 21.50 & 21.74 \\
   2017-08-27 & 20.36 & 20.75 & 21.00 \\
      \hline
    \end{tabular}}\label{tab:limmag}
\end{table}

We started HSC observation from Aug 18.23, 2017 (UTC), corresponding to
$0.7$~days after the GW detection, and also performed HSC
observation on Aug 19, 25, and 27. All the observations were carried out
in the $z$-band. The poor visibility of
\GW\ from Maunakea compels us to conduct the survey during
the astronomical twilight. The observations on Aug 25 and 27
concentrate on one field because the target fields set immediately
after the sunset. The survey pointings are selected from HEALPix
grid with resolution of NSIDE=64 by following criteria: higher
probability of \GW\ sky localization and larger number of nearby galaxies
in the GLADE
catalog\footnote{http://aquarius.elte.hu/glade/index.html}
(Table~\ref{tab:pointing}). We also choose the pointings located in
footprints of Pan-STARRS1 (PS1, \cite{ps1survey}) and use the PS1 catalog
and images for astrometric calibration and image subtraction,
respectively. As the fields with smaller right ascension and declination
set earlier, we conduct the observations in order of reaching the
elevation limit earlier. The observed area is 28.9~deg$^2$
corresponding to the $66.0\%$ credible region of \GW\ (Figure~\ref{fig:pointing}).
Exposures used in the following analysis are listed in Table~\ref{tab:exposure}. 

The data are analyzed with {\it hscPipe} v4.0.5, which is a
standard reduction pipeline of HSC \citep{bos17}. It provides full
packages for data analyses of images obtained with HSC, including
bias subtraction, flat fielding, astrometry, flux calibration,
mosaicing, warping, stacking, image subtraction, source detection, and
source measurement. The astrometric and photometric calibration is made
relative to the PS1 catalog with a 4.0~arcsec (24~pixel)
aperture diameter.
Further, in order to select variable sources, we perform image subtraction
between the HSC and archival PS1 $z$-band images using a
package in {\it hscPipe} based on an algorithm proposed by
\citet{ala98}. The PS1 images are adopted as the reference images and
convolved to fit the point spread function (PSF) shape of the HSC images.

\begin{figure}
 \begin{center}
  \includegraphics[width=8cm]{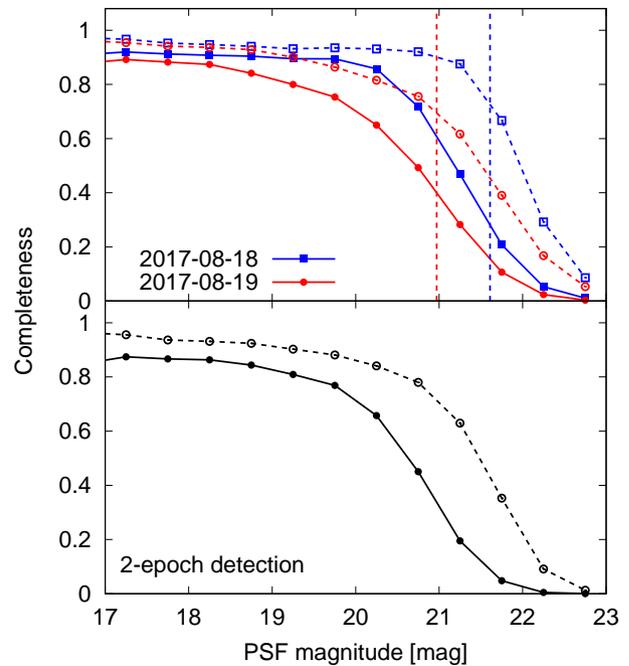} 
 \end{center}
\caption{Completeness of transient detection in the difference images on Aug
 18 (squares) and Aug 19 (circles) (top) and in both of the difference images (bottom). The dashed
 and solid lines represent completeness before and after the
 candidate selection, respectively. The vertical dashed lines show the median
 of $5\sigma$ limiting magnitude before the candidate selection. }\label{fig:completeness}
\end{figure}

We measure the FWHM sizes of PSF in the stacked images with {\it
hscPipe}. These scatters in a wide range from $0.7$ to $1.8$~arcsec depending
on the pointings, especially on the elevation, and the median is
$\sim1.2$~arcsec (Figure~\ref{fig:seeing}). The PSF size statistics is summarized in
Table~\ref{tab:seeing}. The median FWHM size is slightly worse than that
of the image quality of the PS1 $3\pi$ survey \citep{mag16seeing}, and the PSF convolution of
the PS1 image for the image subtraction works well.

After the image subtraction, the $5\sigma$ limiting magnitudes in the difference images are estimated
by measuring standard deviations of fluxes in randomly distributed apertures with a diameter
of twice the FWHM of PSF, and scatter from
$18.3$~mag to $22.5$~mag with a median of $21.3$~mag
(Figure~\ref{fig:limmag} and Table~\ref{tab:limmag}). The $5\sigma$ limiting
magnitudes are mainly determined by the depths of HSC images which are
typically shallower
than those of the PS1 image. In particular, the depths in the pointings observed early on Aug 19
are quite shallow.
We also evaluate completeness of detection by a random injection and 
detection of artificial point sources with various 
magnitude (dashed lines in Figure~\ref{fig:completeness}). The magnitude
of artificial point sources are fixed in time. The large
diversity in the depth of images taken on Aug 19 causes the shallow
dependence of completeness on the PSF magnitude of artificial
sources. The median of $5\sigma$ limiting magnitude is roughly
comparable to the $70\%$ completeness magnitude.

As the detected sources include many bogus detection, candidate
selection is performed as done in \citet{uts17gw151226}. Criteria for
the detection in a single difference image are
(1) $| (S/N)_{\rm PSF}|>5$,
(2) $(b/a) / (b/a)_{\rm PSF} > 0.65$ where $a$ and $b$ are the lengths
of the major and minor axes of a shape of a source, respectively,
(3) $0.7<{\rm FWHM}/{\rm (FWHM)}_{\rm PSF} < 1.3$,
and (4) PSF-subtracted residual $<3\sigma$.
These criteria confirm a high confidence level of
detection and a stellar-like shape of a source.
Further, we impose the sources to be detected in both of the difference
images on Aug 18 and 19, and find 1551 sources.
We also evaluate the completeness of this candidate selection with the
artificial point sources (solid lines in
Figure~\ref{fig:completeness}). The candidate selection makes the
$50\%$ completeness magnitudes shallower by $0.7-0.8$~mag. The completeness of
the two-epoch detection is comparable to that on Aug 19 because the
observation on Aug 19 is shallower than that on Aug 18. The $50\%$
completeness magnitude for two-epoch detection is $20.6$~mag.

The two-epoch detection is only possible for the fields with the
archival PS1
images and the HSC images on both of Aug 18 and 19. The resultant area for the transient
search is 23.6~deg$^2$ corresponding to the $56.6\%$ credible region of
\GW.

\section{Transient search and characteristics}

\subsection{Source screening}

\begin{figure}
 \begin{center}
  \includegraphics[width=8cm]{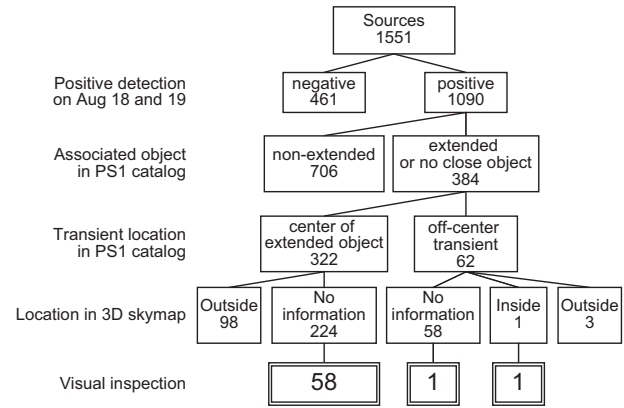} 
 \end{center}
\caption{Flowchart of the candidate screening process. The number in each box
 represents the number of remaining sources after each screening. }\label{fig:flowchart}
\end{figure}

Since the 1551 sources include sources unrelated to \GW, we need to
screen them in order to pick up candidates that may be related to
\GW. We adopt a procedure shown in the flowchart
(Figure~\ref{fig:flowchart}).

\begin{figure}
 \begin{center}
  \includegraphics[width=8cm]{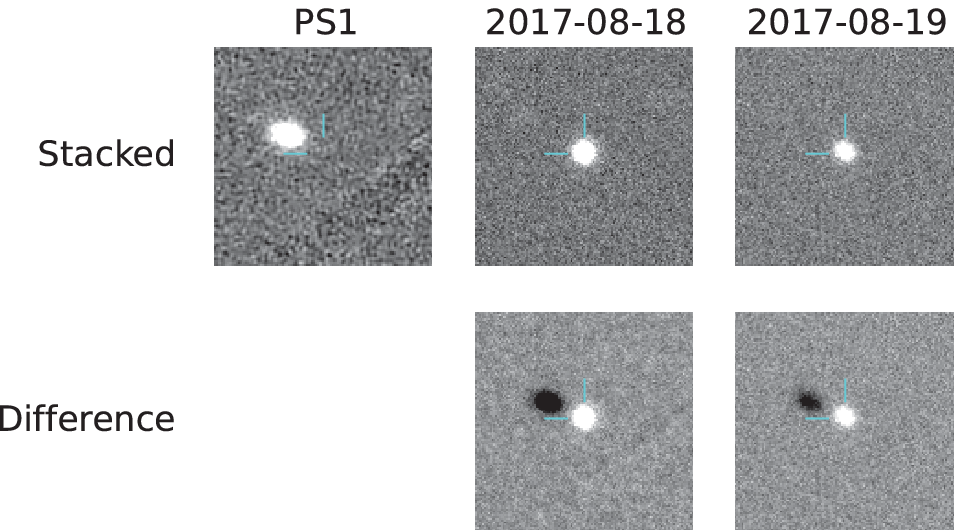} 
  \includegraphics[width=8cm]{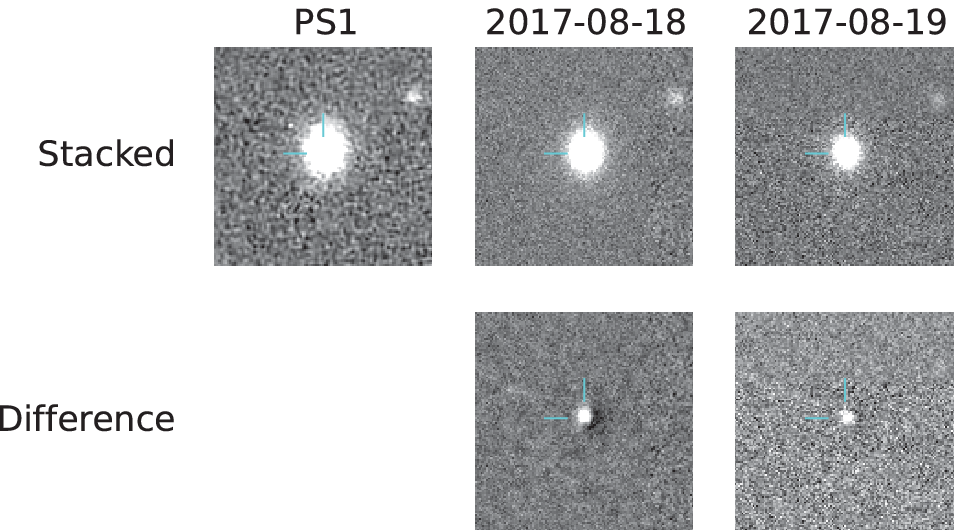}
 \end{center}
\caption{Example of sources excluded by the visual inspection: (Top)
 high proper motion stars and (bottom) bogus detection at the center of
 the extended objects. The lengths of ticks
 are $2$~arcsec and the figure size is $20\times20$~arcsec$^2$.}\label{fig:bogus}
\end{figure}

\begin{figure*}
 \begin{center}
  \includegraphics[width=16cm]{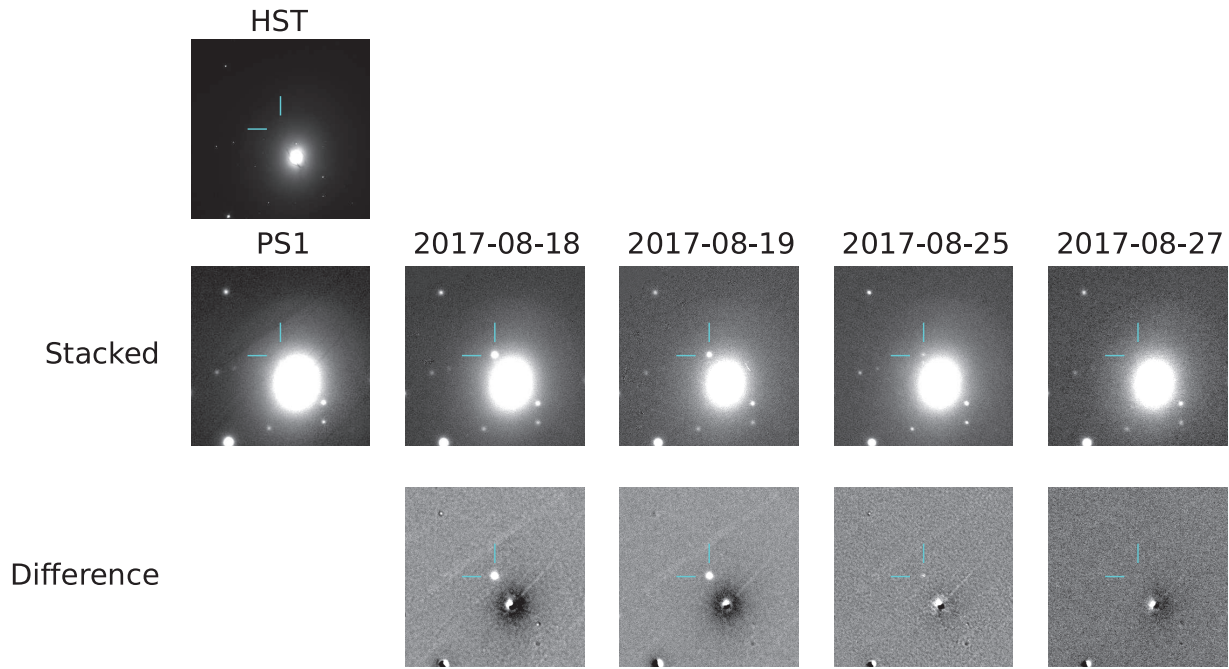} 
 \end{center}
\caption{Stacked and difference $z$-band images of \NameA\ (a.k.a. SSS17a/DLT17ck) associated with
 NGC~4993 located in the 3D skymap of \GW. The archival HST ACS image is also shown. The lengths of ticks
 are $11$~arcsec and the figure size is $56\times56$~arcsec$^2$. }\label{fig:images}
\end{figure*}

First of all, the flux of optical counterpart of \GW\ needs not to be
negative on Aug 18 and 19. We exclude sources having 
significantly negative fluxes ($<-3\sigma$) on Aug 18 or 19. We also
rule out sources 
associated with stellar-like objects in the PS1 catalog
\citep{mag16,fle16}\footnote{https://panstarrs.stsci.edu/} with a
separation of
$<1.0$~arcsec. Here we adopt the larger separation, similar to the
typical seeing size, than the astrometric
error in order to remove bogus detection that frequently appears around a bright star. According
to the number density of stellar-like objects in the PS1 catalog, this
exclusion reduces only $0.2\%$ of the survey fields. After these
screening, 384 sources remain. While 322 sources are located at the
center of extended objects in the PS1 catalog, 62 sources have
separations with $>1.0$~arcsec to any objects in the PS1 catalog.

We further exclude sources associated with PS1 objects that is firmly
located outside of the 3D skymap derived from the GW observations
(LALInference\_v2.fits.gz;
\cite{GW170817LVCpaper}), adopting the GLADE v2 catalog and NASA/IPAC
Extragalactic Database (NED)\footnote{https://ned.ipac.caltech.edu/}.
While we primarily employ the distance in the GLADE catalog, we replace
it with the redshift-independent distance in NED if the associated PS1 objects or one of a
galaxy pair containing the associated PS1 objects have information
\citep{tul88,wil97,fre01,the07,sor14,spr14}, and with the
redshift-dependent distance in NED \citep{mou00} if no distance information is
available in the GLADE catalog.
We search for possibly associated galaxies in the GLADE catalog or NED
with a separation of $<2.0$~arcsec for the 322
sources at the center of extended PS1 objects, which is smaller than the
criteria to identify duplicate galaxies in the GLADE catalog
($3.6$~arcsec), and with a separation of
$<15.0$~arcsec for the 62 off-center sources, which corresponds to
a separation of $<3$~kpc at a distance of $40$~Mpc.
If the 3D probability of \GW\ occurrence at the location and distance of
the associated PS1 object with a HEALPix 3D grid with resolution of
NSIDE=1024 is less than $10^{-3}$ of the maximum probability, the source
is ruled out. This screening reduces the number of sources to 224
sources at the center of extended PS1 objects and 59 off-center sources.

There is only one source (\NameA)
associated with a PS1 object that is
located in the 3D skymap. The detail of \NameA\
is described in
the next subsection. On the other hand, the other 282 sources do
not have distance measurement in the GLADE catalog or NED. 
After the catalog matching, four of the authors remove bogus and
high proper motion stars by visual inspection
(Figure~\ref{fig:bogus}). The number of final candidates,
that may be related to \GW, is 60 (Table~\ref{tab:candidate}). We note that 58
candidates are located at the center of extended PS1 objects and that
some of them could be active galactic nuclei (AGN) or indistinguishable residuals resulting from different instrumental
signatures between PS1 and HSC, but we conservatively hold them as candidates.

\begin{figure}
 \begin{center}
  \includegraphics[width=8cm]{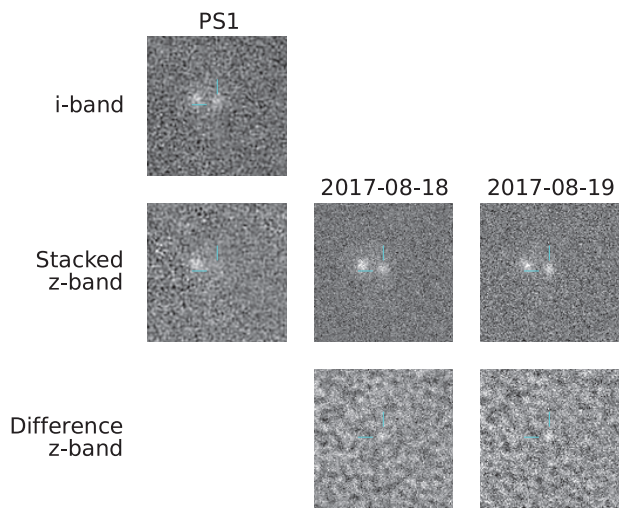} 
 \end{center}
\caption{Stacked and difference $z$-band images of an off-center candidate
 \NameC. The archival PS1 $i$-band image is also shown. The lengths of ticks
 are $2$~arcsec and the figure size is $20\times20$~arcsec$^2$.}\label{fig:imagesC}
\end{figure}

\begin{longtable}{ccc}
 \caption{60 final candidates.}\label{tab:candidate}
 \hline
   Name & R.A. & decl. \\
   & (J2000) & (J2000)\\ 
 \hline
 \endfirsthead
 \hline
   Name & R.A. & decl. \\
   & (J2000) & (J2000)\\ 
 \hline
 \endhead
 \endfoot
 \multicolumn{3}{l}{Some of the candidates at the center of the extended objects}\\
 \multicolumn{3}{l}{could be AGN or indistinguishable residuals resulting from} \\
 \multicolumn{3}{l}{different instrumental signatures between PS1 and HSC.} \\
 \endlastfoot
   \multicolumn{3}{c}{Off-center candidates}\\ \hline
J-GEM17bog & $13^{h}04^{m}44.\hspace{-.55ex}^s11$ & $-22^\circ37'07.\hspace{-.5ex}''2$ \\
J-GEM17btc & $13^{h}09^{m}48.\hspace{-.55ex}^s07$ & $-23^\circ22'53.\hspace{-.5ex}''4$ \\ \hline
   \multicolumn{3}{c}{Candidates at the center of extended objects}\\ \hline
J-GEM17adx & $13^{h}17^{m}42.\hspace{-.55ex}^s18$ & $-27^\circ49'20.\hspace{-.5ex}''7$ \\
J-GEM17aiu & $13^{h}21^{m}26.\hspace{-.55ex}^s97$ & $-27^\circ38'13.\hspace{-.5ex}''5$ \\
J-GEM17aoh & $13^{h}18^{m}25.\hspace{-.55ex}^s05$ & $-25^\circ34'35.\hspace{-.5ex}''1$ \\
J-GEM17aop & $13^{h}17^{m}12.\hspace{-.55ex}^s45$ & $-26^\circ35'21.\hspace{-.5ex}''5$ \\
J-GEM17apm & $13^{h}16^{m}07.\hspace{-.55ex}^s29$ & $-26^\circ00'13.\hspace{-.5ex}''8$ \\
J-GEM17aqg & $13^{h}15^{m}37.\hspace{-.55ex}^s92$ & $-26^\circ08'51.\hspace{-.5ex}''4$ \\
J-GEM17aqh & $13^{h}15^{m}32.\hspace{-.55ex}^s63$ & $-25^\circ59'03.\hspace{-.5ex}''2$ \\
J-GEM17aqk & $13^{h}15^{m}33.\hspace{-.55ex}^s44$ & $-25^\circ43'25.\hspace{-.5ex}''4$ \\
J-GEM17auc & $13^{h}12^{m}56.\hspace{-.55ex}^s78$ & $-25^\circ53'12.\hspace{-.5ex}''4$ \\
J-GEM17avc & $13^{h}11^{m}55.\hspace{-.55ex}^s08$ & $-25^\circ33'48.\hspace{-.5ex}''2$ \\
J-GEM17aws & $13^{h}08^{m}15.\hspace{-.55ex}^s94$ & $-24^\circ05'35.\hspace{-.5ex}''3$ \\
J-GEM17axt & $13^{h}07^{m}31.\hspace{-.55ex}^s91$ & $-24^\circ04'21.\hspace{-.5ex}''9$ \\
J-GEM17azj & $13^{h}04^{m}20.\hspace{-.55ex}^s97$ & $-24^\circ05'18.\hspace{-.5ex}''4$ \\
J-GEM17azl & $13^{h}04^{m}26.\hspace{-.55ex}^s30$ & $-24^\circ04'19.\hspace{-.5ex}''2$ \\
J-GEM17bco & $13^{h}12^{m}14.\hspace{-.55ex}^s45$ & $-24^\circ16'53.\hspace{-.5ex}''7$ \\
J-GEM17bek & $13^{h}10^{m}08.\hspace{-.55ex}^s05$ & $-23^\circ58'15.\hspace{-.5ex}''6$ \\
J-GEM17bfi & $13^{h}09^{m}53.\hspace{-.55ex}^s68$ & $-23^\circ49'30.\hspace{-.5ex}''1$ \\
J-GEM17bfs & $13^{h}08^{m}41.\hspace{-.55ex}^s16$ & $-24^\circ38'41.\hspace{-.5ex}''9$ \\
J-GEM17bgk & $13^{h}08^{m}37.\hspace{-.55ex}^s03$ & $-23^\circ57'57.\hspace{-.5ex}''1$ \\
J-GEM17bjh & $13^{h}00^{m}16.\hspace{-.55ex}^s29$ & $-22^\circ43'30.\hspace{-.5ex}''2$ \\
J-GEM17bka & $13^{h}07^{m}47.\hspace{-.55ex}^s41$ & $-23^\circ34'49.\hspace{-.5ex}''3$ \\
J-GEM17ble & $13^{h}07^{m}22.\hspace{-.55ex}^s90$ & $-22^\circ19'41.\hspace{-.5ex}''9$ \\
J-GEM17blv & $13^{h}06^{m}19.\hspace{-.55ex}^s03$ & $-23^\circ01'44.\hspace{-.5ex}''5$ \\
J-GEM17bna & $13^{h}05^{m}31.\hspace{-.55ex}^s84$ & $-22^\circ37'31.\hspace{-.5ex}''5$ \\
J-GEM17bnb & $13^{h}06^{m}12.\hspace{-.55ex}^s02$ & $-22^\circ36'51.\hspace{-.5ex}''6$ \\
J-GEM17bsc & $13^{h}11^{m}55.\hspace{-.55ex}^s67$ & $-23^\circ40'02.\hspace{-.5ex}''1$ \\
J-GEM17bsf & $13^{h}11^{m}21.\hspace{-.55ex}^s40$ & $-22^\circ44'51.\hspace{-.5ex}''5$ \\
J-GEM17bsm & $13^{h}10^{m}51.\hspace{-.55ex}^s41$ & $-23^\circ10'50.\hspace{-.5ex}''3$ \\
J-GEM17bsn & $13^{h}10^{m}24.\hspace{-.55ex}^s42$ & $-23^\circ09'35.\hspace{-.5ex}''7$ \\
J-GEM17bti & $13^{h}09^{m}41.\hspace{-.55ex}^s65$ & $-23^\circ16'04.\hspace{-.5ex}''4$ \\
J-GEM17bvu & $13^{h}08^{m}02.\hspace{-.55ex}^s66$ & $-23^\circ25'52.\hspace{-.5ex}''0$ \\
J-GEM17bvv & $13^{h}08^{m}25.\hspace{-.55ex}^s91$ & $-23^\circ25'10.\hspace{-.5ex}''2$ \\
J-GEM17bvw & $13^{h}08^{m}30.\hspace{-.55ex}^s96$ & $-23^\circ22'46.\hspace{-.5ex}''7$ \\
J-GEM17byn & $13^{h}02^{m}23.\hspace{-.55ex}^s21$ & $-20^\circ50'55.\hspace{-.5ex}''8$ \\
J-GEM17bzt & $12^{h}58^{m}43.\hspace{-.55ex}^s75$ & $-21^\circ12'45.\hspace{-.5ex}''5$ \\
J-GEM17cao & $13^{h}09^{m}44.\hspace{-.55ex}^s22$ & $-22^\circ08'23.\hspace{-.5ex}''6$ \\
J-GEM17cch & $13^{h}04^{m}19.\hspace{-.55ex}^s39$ & $-21^\circ40'18.\hspace{-.5ex}''4$ \\
J-GEM17cea & $12^{h}54^{m}39.\hspace{-.55ex}^s55$ & $-19^\circ20'55.\hspace{-.5ex}''9$ \\
J-GEM17ceh & $13^{h}03^{m}08.\hspace{-.55ex}^s26$ & $-19^\circ44'17.\hspace{-.5ex}''1$ \\
J-GEM17ceo & $13^{h}01^{m}48.\hspace{-.55ex}^s07$ & $-20^\circ33'48.\hspace{-.5ex}''3$ \\
J-GEM17cet & $13^{h}01^{m}58.\hspace{-.55ex}^s68$ & $-19^\circ28'36.\hspace{-.5ex}''2$ \\
J-GEM17cfe & $13^{h}01^{m}15.\hspace{-.55ex}^s18$ & $-19^\circ50'35.\hspace{-.5ex}''4$ \\
J-GEM17cfi & $13^{h}01^{m}07.\hspace{-.55ex}^s91$ & $-19^\circ29'00.\hspace{-.5ex}''7$ \\
J-GEM17cfm & $13^{h}00^{m}44.\hspace{-.55ex}^s29$ & $-20^\circ34'49.\hspace{-.5ex}''4$ \\
J-GEM17cfy & $12^{h}59^{m}29.\hspace{-.55ex}^s55$ & $-20^\circ52'43.\hspace{-.5ex}''2$ \\
J-GEM17cgi & $12^{h}59^{m}43.\hspace{-.55ex}^s16$ & $-19^\circ32'51.\hspace{-.5ex}''9$ \\
J-GEM17cgq & $12^{h}59^{m}03.\hspace{-.55ex}^s38$ & $-20^\circ00'20.\hspace{-.5ex}''9$ \\
J-GEM17cgv & $12^{h}57^{m}47.\hspace{-.55ex}^s33$ & $-20^\circ34'09.\hspace{-.5ex}''6$ \\
J-GEM17cio & $12^{h}56^{m}22.\hspace{-.55ex}^s01$ & $-19^\circ22'38.\hspace{-.5ex}''3$ \\
J-GEM17ciw & $12^{h}55^{m}49.\hspace{-.55ex}^s13$ & $-18^\circ49'01.\hspace{-.5ex}''0$ \\
J-GEM17ciy & $12^{h}55^{m}45.\hspace{-.55ex}^s53$ & $-18^\circ33'42.\hspace{-.5ex}''5$ \\
J-GEM17cjm & $12^{h}55^{m}35.\hspace{-.55ex}^s31$ & $-18^\circ20'19.\hspace{-.5ex}''5$ \\
J-GEM17ckf & $12^{h}54^{m}21.\hspace{-.55ex}^s62$ & $-18^\circ59'05.\hspace{-.5ex}''8$ \\
J-GEM17ckt & $13^{h}01^{m}47.\hspace{-.55ex}^s22$ & $-19^\circ22'23.\hspace{-.5ex}''7$ \\
J-GEM17ckv & $12^{h}58^{m}59.\hspace{-.55ex}^s95$ & $-19^\circ11'28.\hspace{-.5ex}''9$ \\
J-GEM17cld & $12^{h}58^{m}17.\hspace{-.55ex}^s32$ & $-18^\circ39'20.\hspace{-.5ex}''4$ \\
J-GEM17clo & $12^{h}57^{m}59.\hspace{-.55ex}^s20$ & $-19^\circ11'33.\hspace{-.5ex}''0$ \\
J-GEM17clp & $12^{h}57^{m}56.\hspace{-.55ex}^s99$ & $-19^\circ11'14.\hspace{-.5ex}''6$ \\
 \hline
\end{longtable}

\subsection{Properties of candidates}

We investigate properties of remaining 60
candidates.

Figure~\ref{fig:images} shows the candidate with the associated
PS1 object within the 3D skymap of \GW. \NameA\ is located at 
${\rm R.A.=13^{h}09^m48.\hspace{-.55ex}^s07}$, ${\rm decl.=-23^\circ22'53.\hspace{-.5ex}''4}$ (J2000.0),
which is SSS17a/DLT17ck reported by \citet{coulter17,coulter17_paper,valenti17}.
The nearest object in the PS1 catalog is PSO~J130947.744-232257.366
at ${\rm R.A.=13^{h}09^m47.\hspace{-.55ex}^s74}$, ${\rm decl.=-23^\circ22'57.\hspace{-.5ex}''4}$ (J2000.0)
with a separation of $6.0$~arcsec to \NameA,
which is superposed on NGC~4993 and located at $4.6$~arcsec north of
the center of NGC~4993. According to an archival Hubble Space Telescope
(HST) ACS image \citep{bel17}, the PSF shape of PSO~J130947.744-232257.366 is
consistent with stellar-like sources surrounding it and
PSO~J130947.744-232257.366 is unlikely to be relevant to \NameA. Thus, we
conclude that the second closest object NGC~4993 located $10.0$~arcsec
away from \NameA, well within the separation criterion of $15.0$~arcsec,
is associated with \NameA. NGC~4993 is an S0
galaxy at the distance of $\sim40$~Mpc \citep{fre01}. 

Among remaining 59 candidates, one candidate (\NameC) at
${\rm R.A.=13^{h}04^m44.\hspace{-.55ex}^s11}$, ${\rm decl.=-22^\circ37'07.\hspace{-.5ex}''2}$ (J2000.0)
is registered as an
off-center transient (Figure~\ref{fig:imagesC}). However, we marginally find a 
persistent object overlapping with a galaxy in the archival PS1 $i$-band image, which is not
registered in the PS1 catalog.
The other 58 candidates are located at the center of extended PS1
objects. Among them, two candidates are associated with X-ray sources in
the ROSAT catalog \citep{ROSATcatalog} with separations of $15.6$ and
$16.2$~arcsec, and two other candidates are associated with radio
sources in the NRAO VLA Sky Survey (NVSS, $1.4$~GHz) catalog \citep{NVSScatalog} with separations of $0.9$
and $3.0$~arcsec. These four candidates could be AGN showing optical
variability. We also
check the 3XMM-DR7 catalog \citep{XMMcatalog} but there are no
associated sources in the 3XMM-DR7 catalog.
Although some of them might have little possibility of the association
with \GW, we cannot exclude them from candidates of the optical counterpart of \GW.

\begin{figure}
 \begin{center}
  \includegraphics[width=8cm]{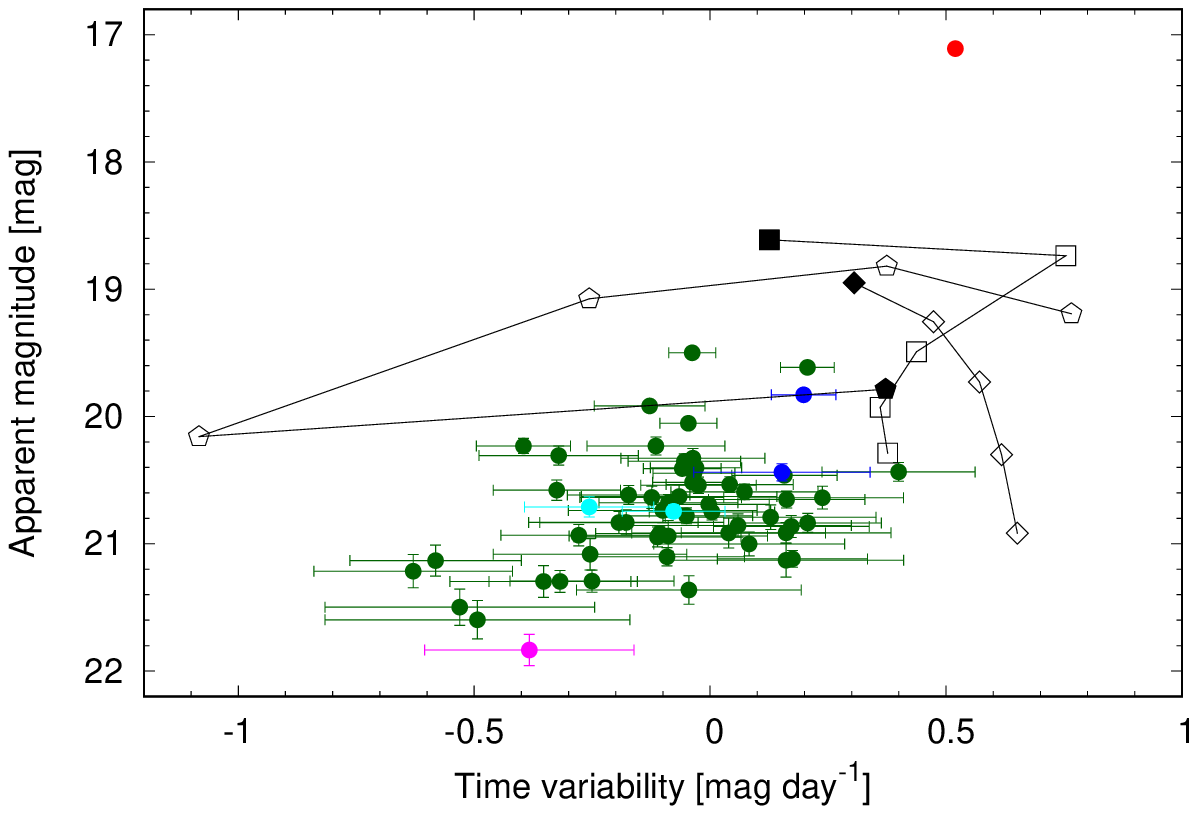} 
 \end{center}
\caption{Magnitude and time variability of the 60 candidates (points with
 error bars, red: \NameA, magenta: \NameC, blue:
 candidates with ROSAT detection, cyan:
 candidates with NVSS detection, and green: the others) and the theoretical
 kilonova models with an ejecta mass of $0.01\Msun$ at $40$~Mpc (points connected with lines,
 \cite{tanaka14,tanaka17gw170817}). The
 models represent dynamical ejecta (diamond),
 and wind with $\ye=0.25$ (square) and $0.3$ (pentagon). Filled marks
 correspond to 1 day after the merger and open marks 
 correspond to 2, 3, 4, and 5 days after the merger. }\label{fig:magdmdt}
\end{figure}

The apparent magnitude on Aug 18 and time variability between Aug 18 and 19 of the
60
candidates are shown in Figure~\ref{fig:magdmdt}.
It is remarkable that \NameA\ is much brighter than the other candidates
with $\sim19-22$~mag. Although this diagram illustrates the
distinguished feature of \NameA, this fact alone is not conclusive
evidence that \NameA\ is the most likely counterpart of \GW.

For comparison, theoretical models of
kilonovae at 1-5 days after the merger are also plotted in Figure~\ref{fig:magdmdt}
\citep{tanaka14,tanaka17gw170817}. Although the observation takes place
at 1 day after the merger, we adopt a wide range of time after the
merger in order to take into account theoretical uncertainties. Although
the rapid time evolution is believed to be a
clue for identification of a kilonova, the time variability at early epochs of the
theoretical models can take any values between $-1.0$ and $+1.0$~mag~day$^{-1}$, which is
consistent with the properties of all the candidates. On the other hand,
there is a discrepancy in the apparent magnitude. However, it can be
explained by different ejecta masses, which could result from 
uncertainties of the equation of state and different efficiency of
viscus heating (\eg \cite{shi17}). Therefore, we cannot
rule out the possibility of any candidates as an EM counterpart of \GW\
from the time variability and the brightness at 1~day after \GW.

Since the distance to the candidates are unknown except for \NameA, we
evaluate the probability $P_{\rm 3D}$ that the associated PS1 object is located
inside of the 3D skymap of \GW, with a luminosity function of galaxies
at a rest wavelength $\lambda$, $\phi(\lambda,M)$,
derived from rest-frame $UBVRI$ luminosity functions \citep{ilb05} and the
{\it Planck} cosmology \citep{planck13} as follows:
\begin{equation}
 P_{\rm 3D}(\lambda_j,m_j) = {\int_{D_{\rm mean}-3\sigma_D}^{D_{\rm mean}+3\sigma_D}
  \phi(\lambda[\lambda_j,D],M[m_j,D]) A(D) dD
  \over{\int_0^\infty} \phi(\lambda[\lambda_j,D],M[m_j,D]) A(D) dD},
  \label{eq:p}
\end{equation}
where $D_{\rm mean}$ and $\sigma_D$ are the mean and standard
deviation of the distance to \GW\ at the position, respectively, $M(m_j,D)$ is the absolute
magnitude of a galaxy with observer-frame $j$-band apparent magnitude $m_j$ at a distance of $D$,
$\lambda(\lambda_j,D)$ is the rest wavelength redshifted from the observed
wavelength $\lambda_j$ with a distance of $D$,
and $A(D)$ is the surface area of observed region at a distance of
$D$. 

We evaluated $P_{\rm 3D}$ for the PS1 objects associated with the
59 candidates using the $r$- and/or $i$-band Kron magnitude in the PS1
catalog. 
We also estimate the probability of
NGC~4993 ($R=12.09$~mag, \cite{lau89}) associated with \NameA, which is $P_{\rm 3D}=64\%$. On the
other hand, the probability of \NameC\ is $9.3\times10^{-3}\%$ and the
probabilities of the other 58 candidates range from $1.2\times10^{-2}\%$
to $2.1\times10^{-1}\%$. Furthermore, the possibility, that
at least one of the 59 candidates including \NameC\ is located in the 3D
skymap of \GW, is only $3.2\%$. Therefore, we conclude that
\NameA\ is more likely, by more than an order of magnitude, to be the
optical counterpart of \GW\ than the other candidates.
The large difference between \NameA\ and the other 59 candidates stems from
the faintness of the associated objects of the other 59 candidates,
which prevent them from being registered in the GLADE catalog or
NED. Given the luminosity function of galaxy and the comoving volume,
the faint objects are likely to be distant objects and thus $P_{\rm 3D}$ of
them are small. We note that the integrand of the denominator in
equation (\ref{eq:p}) is nearly
zero at a redshift of $z\geq0.7$ for all of the 60 candidates and that
these results are almost independent on the adopted filters.

\section{Conclusions}

We have performed the survey for the optical counterpart of \GW\
with Subaru/HSC. Our untargeted transient search covers 23.6~deg$^2$ corresponding
to the $56.6\%$ credible region of \GW\ and reaches the $50\%$ completeness
magnitude of $20.6$~mag. We find 1551 sources with two-epoch detection,
and screen them with the catalog matching and the visual inspection. The
number of our final candidates is 60.

We find only one candidate
\NameA\ with an associated object firmly located within the 3D skymap of \GW. On the other
hand, the other 59 candidates do not have distance information of associated
objects.
The candidates include one off-center candidate other than \NameA,
but it is associated with the marginally-detected persistent 
object in the archival PS1 $i$-band image. The other 58 candidates are located at
the center of extended PS1 objects and could be AGN. Four of them are
actually associated with the ROSAT X-ray sources or NVSS radio sources. However, we can not
rule out the other 59 candidates from our
observations because the kilonova model can have any time variability of
$-1.0$ to $+1.0$~mag~day$^{-1}$ at the early epochs.

Hence, we evaluate the probability that the PS1 object associated
with the candidate is located inside of the 3D skymap of \GW. The
probability of NGC~4993 associated with \NameA\ is
$64\%$, while the possibility, that at least one of the other 59
candidates is located in the 3D skymap, is only $3.2\%$. Therefore, we conclude that
\NameA\ (a.k.a. SSS17a/DLT17ck) is the most-likely and distinguished
candidate as the optical counterpart of \GW. The same conclusion is
brought by the other untargeted wide-field survey with the Dark Energy
Camera (DECam, \cite{GW170817DECampaper}). We note that \NameA\ is intensively
observed by many telescopes, satellites, and instruments (e.g.,
\cite{GW170817MMApaper,utsumi17}).


\begin{ack}
 We thank Dr. Daigo Tomono for his technical comment on operating the
 Subaru telescope and HSC under severe condition and Dr. Nobuhiro Okabe who provided a
 computational resource. This work was supported by MEXT KAKENHI
 (JP17H06363, JP15H00788, JP24103003, JP10147214, JP10147207)
 and JSPS KAKENHI (JP16H02183, JP15H02075, JP26800103,
 JP25800103), the research grant program of the Toyota Foundation
 (D11-R-0830), the natural science grant of the Mitsubishi Foundation,
 the research grant of the Yamada Science Foundation, the NINS program
 for
 cross-disciplinary science study, Inoue Science Research Award from
 Inoue Foundation for Science, Optical \& Near-Infrared Astronomy
 Inter-University Cooperation Program from the MEXT, the National
 Research
 Foundation of South Africa.
\end{ack}

\bibliographystyle{myaasjournal} 
\bibliography{ms}

\begin{thebibliography}{}
\expandafter\ifx\csname natexlab\endcsname\relax\def\natexlab#1{#1}\fi
\providecommand{\url}[1]{\href{#1}{#1}}

\bibitem[{{Abbott} {et~al.}(2016{\natexlab{a}}){Abbott}, {Abbott}, {Abbott},
  {Abernathy}, {Acernese}, {Ackley}, {Adams}, {Adams}, {Addesso}, {Adhikari},
  \& et~al.}]{gw150914}
{Abbott}, B.~P., {et~al.} 2016{\natexlab{a}}, Physical Review Letters, 116,
  061102

\bibitem[{{Abbott} {et~al.}(2016{\natexlab{b}}){Abbott}, {Abbott}, {Abbott},
  {Abernathy}, {Acernese}, {Ackley}, {Adams}, {Adams}, {Addesso}, {Adhikari},
  \& et~al.}]{gw151226}
---. 2016{\natexlab{b}}, Physical Review Letters, 116, 241103

\bibitem[{{Abbott} {et~al.}(2016{\natexlab{c}}){Abbott}, {Abbott}, {Abbott},
  {Abernathy}, {Acernese}, {Ackley}, {Adams}, {Adams}, {Addesso}, {Adhikari},
  \& et~al.}]{lvt151012}
---. 2016{\natexlab{c}}, Physical Review X, 6, 041015

\bibitem[{{Abbott} {et~al.}(2017{\natexlab{a}}){Abbott}, {Abbott}, {Abbott},
  {Acernese}, {Ackley}, {Adams}, {Adams}, {Addesso}, {Adhikari}, {Adya}, \&
  et~al.}]{gw170104}
---. 2017{\natexlab{a}}, Physical Review Letters, 118, 221101

\bibitem[{{Abbott} {et~al.}(2017{\natexlab{b}}){Abbott}, {Abbott}, {Abbott},
  {Acernese}, {Ackley}, {Adams}, {Adams}, {Addesso}, {Adhikari}, {Adya}, \&
  et~al.}]{gw170814}
---. 2017{\natexlab{b}}, Physical Review Letters, 119, 141101

\bibitem[{{Abbott} {et~al.}(2017{\natexlab{c}}){Abbott}, {Abbott}, {Abbott},
  {Acernese}, {Ackley}, {Adams}, {Adams}, {Addesso}, {Adhikari}, {Adya}, \&
  et~al.}]{GW170817LVCpaper}
---. 2017{\natexlab{c}}, Physical Review Letters, 119, 161101

\bibitem[{{Abbott} {et~al.}(2017{\natexlab{d}}){Abbott}, {Abbott}, {Abbott},
  {Acernese}, {Ackley}, {Adams}, {Adams}, {Addesso}, {Adhikari}, {Adya}, \&
  et~al.}]{GW170817MMApaper}
---. 2017{\natexlab{d}}, \apjl, 848, L12

\bibitem[{{Alard} \& {Lupton}(1998)}]{ala98}
{Alard}, C., \& {Lupton}, R.~H. 1998, \apj, 503, 325

\bibitem[{{Barnes} \& {Kasen}(2013)}]{bar13}
{Barnes}, J., \& {Kasen}, D. 2013, \apj, 775, 18

\bibitem[{{Bauswein} {et~al.}(2013){Bauswein}, {Goriely}, \& {Janka}}]{bau13}
{Bauswein}, A., {Goriely}, S., \& {Janka}, H.-T. 2013, \apj, 773, 78

\bibitem[{{Bellini} {et~al.}(2017){Bellini}, {Grogin}, {Hathi}, \&
  {Brown}}]{bel17}
{Bellini}, A., {Grogin}, N.~A., {Hathi}, N., \& {Brown}, T.~M. 2017, {The
  Hubble Space Telescope ``Program of Last Resort''}, Tech. rep.

\bibitem[{{Boller} {et~al.}(2016){Boller}, {Freyberg}, {Tr{\"u}mper}, {Haberl},
  {Voges}, \& {Nandra}}]{ROSATcatalog}
{Boller}, T., {Freyberg}, M.~J., {Tr{\"u}mper}, J., {Haberl}, F., {Voges}, W.,
  \& {Nandra}, K. 2016, \aap, 588, A103

\bibitem[{{Bosch} {et~al.}(2017){Bosch}, {Armstrong}, {Bickerton}, {Furusawa},
  {Ikeda}, {Koike}, {Lupton}, {Mineo}, {Price}, {Takata}, {Tanaka}, {Yasuda},
  {AlSayyad}, {Becker}, {Coulton}, {Coupon}, {Garmilla}, {Huang}, {Krughoff},
  {Lang}, {Leauthaud}, {Lim}, {Lust}, {MacArthur}, {Mandelbaum}, {Miyatake},
  {Miyazaki}, {Murata}, {More}, {Okura}, {Owen}, {Swinbank}, {Strauss},
  {Yamada}, \& {Yamanoi}}]{bos17}
{Bosch}, J., {et~al.} 2017, ArXiv e-prints, arXiv:1705.06766

\bibitem[{{Chambers} {et~al.}(2016){Chambers}, {Magnier}, {Metcalfe},
  {Flewelling}, {Huber}, {Waters}, {Denneau}, {Draper}, {Farrow}, {Finkbeiner},
  {Holmberg}, {Koppenhoefer}, {Price}, {Saglia}, {Schlafly}, {Smartt},
  {Sweeney}, {Wainscoat}, {Burgett}, {Grav}, {Heasley}, {Hodapp}, {Jedicke},
  {Kaiser}, {Kudritzki}, {Luppino}, {Lupton}, {Monet}, {Morgan}, {Onaka},
  {Stubbs}, {Tonry}, {Banados}, {Bell}, {Bender}, {Bernard}, {Botticella},
  {Casertano}, {Chastel}, {Chen}, {Chen}, {Cole}, {Deacon}, {Frenk},
  {Fitzsimmons}, {Gezari}, {Goessl}, {Goggia}, {Goldman}, {Grebel}, {Hambly},
  {Hasinger}, {Heavens}, {Heckman}, {Henderson}, {Henning}, {Holman}, {Hopp},
  {Ip}, {Isani}, {Keyes}, {Koekemoer}, {Kotak}, {Long}, {Lucey}, {Liu},
  {Martin}, {McLean}, {Morganson}, {Murphy}, {Nieto-Santisteban}, {Norberg},
  {Peacock}, {Pier}, {Postman}, {Primak}, {Rae}, {Rest}, {Riess}, {Riffeser},
  {Rix}, {Roser}, {Schilbach}, {Schultz}, {Scolnic}, {Szalay}, {Seitz},
  {Shiao}, {Small}, {Smith}, {Soderblom}, {Taylor}, {Thakar}, {Thiel},
  {Thilker}, {Urata}, {Valenti}, {Walter}, {Watters}, {Werner}, {White},
  {Wood-Vasey}, \& {Wyse}}]{ps1survey}
{Chambers}, K.~C., {et~al.} 2016, ArXiv e-prints, arXiv:1612.05560

\bibitem[{{Condon} {et~al.}(1998){Condon}, {Cotton}, {Greisen}, {Yin},
  {Perley}, {Taylor}, \& {Broderick}}]{NVSScatalog}
{Condon}, J.~J., {Cotton}, W.~D., {Greisen}, E.~W., {Yin}, Q.~F., {Perley},
  R.~A., {Taylor}, G.~B., \& {Broderick}, J.~J. 1998, \aj, 115, 1693

\bibitem[{{Connaughton} {et~al.}(2016){Connaughton}, {Burns}, {Goldstein},
  {Blackburn}, {Briggs}, {Zhang}, {Camp}, {Christensen}, {Hui}, {Jenke},
  {Littenberg}, {McEnery}, {Racusin}, {Shawhan}, {Singer}, {Veitch},
  {Wilson-Hodge}, {Bhat}, {Bissaldi}, {Cleveland}, {Fitzpatrick}, {Giles},
  {Gibby}, {von Kienlin}, {Kippen}, {McBreen}, {Mailyan}, {Meegan}, {Paciesas},
  {Preece}, {Roberts}, {Sparke}, {Stanbro}, {Toelge}, \&
  {Veres}}]{con16GW150914}
{Connaughton}, V., {et~al.} 2016, \apjl, 826, L6

\bibitem[{{Connaughton} {et~al.}(2017){Connaughton}, {Name}, {Name}, {Name},
  {Name}, {Name}, {Name}, {Name}, {Name}, {Name}, \& {Name}}]{GW170817GBM}
---. 2017, GCN Circ., 21506

\bibitem[{{Coulter} {et~al.}(2017{\natexlab{a}}){Coulter}, {Kilpatrick},
  {Siebert}, {Foley}, J., {Drout}, \& {Piro}}]{coulter17}
{Coulter}, D.~A., {Kilpatrick}, C.~D., {Siebert}, M.~R., {Foley}, R.~J., J.,
  S.~B., {Drout}, M. R.~{Simon}, J.~S., \& {Piro}, A.~L. 2017{\natexlab{a}},
  GCN Circ., 21529

\bibitem[{{Coulter} {et~al.}(2017{\natexlab{b}}){Coulter}, {Foley},
  {Kilpatrick}, {Drout}, {Piro}, {Shappee}, {Siebert}, {Simon}, {Ulloa},
  {Kasen}, {Madore}, {Murguia-Berthier}, {Pan}, {Prochaska}, {Ramirez-Ruiz},
  {Rest}, \& {Rojas-Bravo}}]{coulter17_paper}
{Coulter}, D.~A., {et~al.} 2017{\natexlab{b}}, ArXiv e-prints, arXiv:1710.05452

\bibitem[{{Dessart} {et~al.}(2009){Dessart}, {Ott}, {Burrows}, {Rosswog}, \&
  {Livne}}]{des09}
{Dessart}, L., {Ott}, C.~D., {Burrows}, A., {Rosswog}, S., \& {Livne}, E. 2009,
  \apj, 690, 1681

\bibitem[{{Eichler} {et~al.}(1989){Eichler}, {Livio}, {Piran}, \&
  {Schramm}}]{eic89}
{Eichler}, D., {Livio}, M., {Piran}, T., \& {Schramm}, D.~N. 1989, \nat, 340,
  126

\bibitem[{{Fern{\'a}ndez} \& {Metzger}(2013)}]{fer13}
{Fern{\'a}ndez}, R., \& {Metzger}, B.~D. 2013, \mnras, 435, 502

\bibitem[{{Flewelling} {et~al.}(2016){Flewelling}, {Magnier}, {Chambers},
  {Heasley}, {Holmberg}, {Huber}, {Sweeney}, {Waters}, {Chen}, {Farrow},
  {Hasinger}, {Henderson}, {Long}, {Metcalfe}, {Nieto-Santisteban}, {Norberg},
  {Saglia}, {Szalay}, {Rest}, {Thakar}, {Tonry}, {Valenti}, {Werner}, {White},
  {Denneau}, {Draper}, {Hodapp}, {Jedicke}, {Kaiser}, {Kudritzki}, {Price},
  {Wainscoat}, {Chastel}, {McClean}, {Postman}, \& {Shiao}}]{fle16}
{Flewelling}, H.~A., {et~al.} 2016, ArXiv e-prints, arXiv:1612.05243

\bibitem[{{Freedman} {et~al.}(2001){Freedman}, {Madore}, {Gibson}, {Ferrarese},
  {Kelson}, {Sakai}, {Mould}, {Kennicutt}, {Ford}, {Graham}, {Huchra},
  {Hughes}, {Illingworth}, {Macri}, \& {Stetson}}]{fre01}
{Freedman}, W.~L., {et~al.} 2001, \apj, 553, 47

\bibitem[{{Goriely} {et~al.}(2011){Goriely}, {Bauswein}, \& {Janka}}]{gor11}
{Goriely}, S., {Bauswein}, A., \& {Janka}, H.-T. 2011, \apjl, 738, L32

\bibitem[{{Greiner} {et~al.}(2016){Greiner}, {Burgess}, {Savchenko}, \&
  {Yu}}]{gre16GW150914}
{Greiner}, J., {Burgess}, J.~M., {Savchenko}, V., \& {Yu}, H.-F. 2016, \apjl,
  827, L38

\bibitem[{{Hotokezaka} {et~al.}(2013){Hotokezaka}, {Kyutoku}, \&
  {Shibata}}]{hot13}
{Hotokezaka}, K., {Kyutoku}, K., \& {Shibata}, M. 2013, \prd, 87, 044001

\bibitem[{{Hulse} \& {Taylor}(1975)}]{hul75}
{Hulse}, R.~A., \& {Taylor}, J.~H. 1975, \apjl, 195, L51

\bibitem[{{Ilbert} {et~al.}(2005){Ilbert}, {Tresse}, {Zucca}, {Bardelli},
  {Arnouts}, {Zamorani}, {Pozzetti}, {Bottini}, {Garilli}, {Le Brun}, {Le
  F{\`e}vre}, {Maccagni}, {Picat}, {Scaramella}, {Scodeggio}, {Vettolani},
  {Zanichelli}, {Adami}, {Arnaboldi}, {Bolzonella}, {Cappi}, {Charlot},
  {Contini}, {Foucaud}, {Franzetti}, {Gavignaud}, {Guzzo}, {Iovino},
  {McCracken}, {Marano}, {Marinoni}, {Mathez}, {Mazure}, {Meneux}, {Merighi},
  {Paltani}, {Pello}, {Pollo}, {Radovich}, {Bondi}, {Bongiorno}, {Busarello},
  {Ciliegi}, {Lamareille}, {Mellier}, {Merluzzi}, {Ripepi}, \& {Rizzo}}]{ilb05}
{Ilbert}, O., {et~al.} 2005, \aap, 439, 863

\bibitem[{{Kasen} {et~al.}(2013){Kasen}, {Badnell}, \& {Barnes}}]{kas13}
{Kasen}, D., {Badnell}, N.~R., \& {Barnes}, J. 2013, \apj, 774, 25

\bibitem[{{Kasen} {et~al.}(2015){Kasen}, {Fern{\'a}ndez}, \& {Metzger}}]{kas15}
{Kasen}, D., {Fern{\'a}ndez}, R., \& {Metzger}, B.~D. 2015, \mnras, 450, 1777

\bibitem[{{Kasliwal} {et~al.}(2016){Kasliwal}, {Cenko}, {Singer}, {Corsi},
  {Cao}, {Barlow}, {Bhalerao}, {Bellm}, {Cook}, {Duggan}, {Ferretti}, {Frail},
  {Horesh}, {Kendrick}, {Kulkarni}, {Lunnan}, {Palliyaguru}, {Laher}, {Masci},
  {Manulis}, {Miller}, {Nugent}, {Perley}, {Prince}, {Quimby}, {Rana},
  {Rebbapragada}, {Sesar}, {Singhal}, {Surace}, \& {Van Sistine}}]{PTFgw150914}
{Kasliwal}, M.~M., {et~al.} 2016, \apjl, 824, L24

\bibitem[{{Kisaka} {et~al.}(2016){Kisaka}, {Ioka}, \& {Nakar}}]{kis16}
{Kisaka}, S., {Ioka}, K., \& {Nakar}, E. 2016, \apj, 818, 104

\bibitem[{{Korobkin} {et~al.}(2012){Korobkin}, {Rosswog}, {Arcones}, \&
  {Winteler}}]{kor12}
{Korobkin}, O., {Rosswog}, S., {Arcones}, A., \& {Winteler}, C. 2012, \mnras,
  426, 1940

\bibitem[{{Kulkarni}(2005)}]{kul05}
{Kulkarni}, S.~R. 2005, ArXiv Astrophysics e-prints, astro-ph/0510256

\bibitem[{{Lattimer} \& {Schramm}(1974)}]{lat74}
{Lattimer}, J.~M., \& {Schramm}, D.~N. 1974, \apjl, 192, L145

\bibitem[{{Lauberts} \& {Valentijn}(1989)}]{lau89}
{Lauberts}, A., \& {Valentijn}, E.~A. 1989, {The surface photometry catalogue
  of the ESO-Uppsala galaxies}

\bibitem[{{Li} \& {Paczy{\'n}ski}(1998)}]{li98}
{Li}, L.-X., \& {Paczy{\'n}ski}, B. 1998, \apjl, 507, L59

\bibitem[{{Magnier} {et~al.}(2016{\natexlab{a}}){Magnier}, {Schlafly},
  {Finkbeiner}, {Tonry}, {Goldman}, {R{\"o}ser}, {Schilbach}, {Chambers},
  {Flewelling}, {Huber}, {Price}, {Sweeney}, {Waters}, {Denneau}, {Draper},
  {Hodapp}, {Jedicke}, {Kudritzki}, {Metcalfe}, {Stubbs}, \&
  {Wainscoast}}]{mag16seeing}
{Magnier}, E.~A., {et~al.} 2016{\natexlab{a}}, ArXiv e-prints, arXiv:1612.05242

\bibitem[{{Magnier} {et~al.}(2016{\natexlab{b}}){Magnier}, {Sweeney},
  {Chambers}, {Flewelling}, {Huber}, {Price}, {Waters}, {Denneau}, {Draper},
  {Jedicke}, {Hodapp}, {Kaiser}, {Kudritzki}, {Metcalfe}, {Stubbs}, \&
  {Wainscoast}}]{mag16}
---. 2016{\natexlab{b}}, ArXiv e-prints, arXiv:1612.05244

\bibitem[{{Metzger}(2017)}]{met17}
{Metzger}, B.~D. 2017, Living Reviews in Relativity, 20, 3

\bibitem[{{Metzger} \& {Berger}(2012)}]{met12}
{Metzger}, B.~D., \& {Berger}, E. 2012, \apj, 746, 48

\bibitem[{{Metzger} \& {Fern{\'a}ndez}(2014)}]{met14}
{Metzger}, B.~D., \& {Fern{\'a}ndez}, R. 2014, \mnras, 441, 3444

\bibitem[{{Metzger} {et~al.}(2010){Metzger}, {Mart{\'{\i}}nez-Pinedo},
  {Darbha}, {Quataert}, {Arcones}, {Kasen}, {Thomas}, {Nugent}, {Panov}, \&
  {Zinner}}]{met10}
{Metzger}, B.~D., {et~al.} 2010, \mnras, 406, 2650

\bibitem[{{Miyazaki} {et~al.}(2012){Miyazaki}, {Komiyama}, {Nakaya}, {Kamata},
  {Doi}, {Hamana}, {Karoji}, {Furusawa}, {Kawanomoto}, {Morokuma}, {Ishizuka},
  {Nariai}, {Tanaka}, {Uraguchi}, {Utsumi}, {Obuchi}, {Okura}, {Oguri},
  {Takata}, {Tomono}, {Kurakami}, {Namikawa}, {Usuda}, {Yamanoi}, {Terai},
  {Uekiyo}, {Yamada}, {Koike}, {Aihara}, {Fujimori}, {Mineo}, {Miyatake},
  {Yasuda}, {Nishizawa}, {Saito}, {Tanaka}, {Uchida}, {Katayama}, {Wang},
  {Chen}, {Lupton}, {Loomis}, {Bickerton}, {Price}, {Gunn}, {Suzuki},
  {Miyazaki}, {Muramatsu}, {Yamamoto}, {Endo}, {Ezaki}, {Itoh}, {Miwa},
  {Yokota}, {Matsuda}, {Ebinuma}, \& {Takeshi}}]{miy12}
{Miyazaki}, S., {et~al.} 2012, in Society of Photo-Optical Instrumentation
  Engineers (SPIE) Conference Series, Vol. 8446, Society of Photo-Optical
  Instrumentation Engineers (SPIE) Conference Series, 0

\bibitem[{{Morokuma} {et~al.}(2016){Morokuma}, {Tanaka}, {Asakura}, {Abe},
  {Tristram}, {Utsumi}, {Doi}, {Fujisawa}, {Itoh}, {Itoh}, {Kawabata}, {Kawai},
  {Kuroda}, {Matsubayashi}, {Motohara}, {Murata}, {Nagayama}, {Ohta}, {Saito},
  {Tamura}, {Tominaga}, {Uemura}, {Yanagisawa}, {Yatsu}, \&
  {Yoshida}}]{JGEMgw150914}
{Morokuma}, T., {et~al.} 2016, \pasj, 68, L9

\bibitem[{{Mould} {et~al.}(2000){Mould}, {Huchra}, {Freedman}, {Kennicutt},
  {Ferrarese}, {Ford}, {Gibson}, {Graham}, {Hughes}, {Illingworth}, {Kelson},
  {Macri}, {Madore}, {Sakai}, {Sebo}, {Silbermann}, \& {Stetson}}]{mou00}
{Mould}, J.~R., {et~al.} 2000, \apj, 529, 786

\bibitem[{{Planck Collaboration} {et~al.}(2014){Planck Collaboration}, {Ade},
  {Aghanim}, {Armitage-Caplan}, {Arnaud}, {Ashdown}, {Atrio-Barandela},
  {Aumont}, {Baccigalupi}, {Banday}, \& et~al.}]{planck13}
{Planck Collaboration}, {et~al.} 2014, \aap, 571, A16

\bibitem[{{Rosen} {et~al.}(2016){Rosen}, {Webb}, {Watson}, {Ballet}, {Barret},
  {Braito}, {Carrera}, {Ceballos}, {Coriat}, {Della Ceca}, {Denkinson},
  {Esquej}, {Farrell}, {Freyberg}, {Gris{\'e}}, {Guillout}, {Heil},
  {Koliopanos}, {Law-Green}, {Lamer}, {Lin}, {Martino}, {Michel}, {Motch},
  {Nebot Gomez-Moran}, {Page}, {Page}, {Page}, {Pakull}, {Pye}, {Read},
  {Rodriguez}, {Sakano}, {Saxton}, {Schwope}, {Scott}, {Sturm}, {Traulsen},
  {Yershov}, \& {Zolotukhin}}]{XMMcatalog}
{Rosen}, S.~R., {et~al.} 2016, \aap, 590, A1

\bibitem[{{Rosswog} {et~al.}(1999){Rosswog}, {Liebend{\"o}rfer}, {Thielemann},
  {Davies}, {Benz}, \& {Piran}}]{ros99}
{Rosswog}, S., {Liebend{\"o}rfer}, M., {Thielemann}, F.-K., {Davies}, M.~B.,
  {Benz}, W., \& {Piran}, T. 1999, \aap, 341, 499

\bibitem[{{Savchenko} {et~al.}(2016){Savchenko}, {Ferrigno}, {Mereghetti},
  {Natalucci}, {Bazzano}, {Bozzo}, {Brandt}, {Courvoisier}, {Diehl}, {Hanlon},
  {von Kienlin}, {Kuulkers}, {Laurent}, {Lebrun}, {Roques}, {Ubertini}, \&
  {Weidenspointner}}]{sav16GW150914}
{Savchenko}, V., {et~al.} 2016, \apjl, 820, L36

\bibitem[{{Savchenko} {et~al.}(2017{\natexlab{a}}){Savchenko}, {Name}, {Name},
  {Name}, {Name}, {Name}, {Name}, {Name}, {Name}, {Name}, \&
  {Name}}]{GW170817INTEGRAL}
---. 2017{\natexlab{a}}, GCN Circ., 21507

\bibitem[{{Savchenko} {et~al.}(2017{\natexlab{b}}){Savchenko}, {Ferrigno},
  {Kuulkers}, {Bazzano}, {Bozzo}, {Brandt}, {Chenevez}, {Courvoisier}, {Diehl},
  {Domingo}, {Hanlon}, {Jourdain}, {von Kienlin}, {Laurent}, {Lebrun},
  {Lutovinov}, {Martin-Carrillo}, {Mereghetti}, {Natalucci}, {Rodi}, {Roques},
  {Sunyaev}, \& {Ubertini}}]{GW170817INTEGRALpaper}
---. 2017{\natexlab{b}}, \apjl, 848, L15

\bibitem[{{Schlafly} \& {Finkbeiner}(2011)}]{sch11}
{Schlafly}, E.~F., \& {Finkbeiner}, D.~P. 2011, \apj, 737, 103

\bibitem[{{Shibata} {et~al.}(2017){Shibata}, {Kiuchi}, \& {Sekiguchi}}]{shi17}
{Shibata}, M., {Kiuchi}, K., \& {Sekiguchi}, Y.-i. 2017, \prd, 95, 083005

\bibitem[{{Smartt} {et~al.}(2016){Smartt}, {Chambers}, {Smith}, {Huber},
  {Young}, {Cappellaro}, {Wright}, {Coughlin}, {Schultz}, {Denneau},
  {Flewelling}, {Heinze}, {Magnier}, {Primak}, {Rest}, {Sherstyuk}, {Stalder},
  {Stubbs}, {Tonry}, {Waters}, {Willman}, {Anderson}, {Baltay}, {Botticella},
  {Campbell}, {Dennefeld}, {Chen}, {Della Valle}, {Elias-Rosa}, {Fraser},
  {Inserra}, {Kankare}, {Kotak}, {Kupfer}, {Harmanen}, {Galbany}, {Gal-Yam},
  {Le Guillou}, {Lyman}, {Maguire}, {Mitra}, {Nicholl}, {Olivares E},
  {Rabinowitz}, {Razza}, {Sollerman}, {Smith}, {Terreran}, {Valenti}, {Gibson},
  \& {Goggia}}]{PSgw150914}
{Smartt}, S.~J., {et~al.} 2016, \mnras, 462, 4094

\bibitem[{{Soares-Santos} {et~al.}(2016){Soares-Santos}, {Kessler}, {Berger},
  {Annis}, {Brout}, {Buckley-Geer}, {Chen}, {Cowperthwaite}, {Diehl}, {Doctor},
  {Drlica-Wagner}, {Farr}, {Finley}, {Flaugher}, {Foley}, {Frieman}, {Gruendl},
  {Herner}, {Holz}, {Lin}, {Marriner}, {Neilsen}, {Rest}, {Sako}, {Scolnic},
  {Sobreira}, {Walker}, {Wester}, {Yanny}, {Abbott}, {Abdalla}, {Allam},
  {Armstrong}, {Banerji}, {Benoit-L{\'e}vy}, {Bernstein}, {Bertin}, {Brown},
  {Burke}, {Capozzi}, {Carnero Rosell}, {Carrasco Kind}, {Carretero},
  {Castander}, {Cenko}, {Chornock}, {Crocce}, {D'Andrea}, {da Costa}, {Desai},
  {Dietrich}, {Drout}, {Eifler}, {Estrada}, {Evrard}, {Fairhurst}, {Fernandez},
  {Fischer}, {Fong}, {Fosalba}, {Fox}, {Fryer}, {Garcia-Bellido}, {Gaztanaga},
  {Gerdes}, {Goldstein}, {Gruen}, {Gutierrez}, {Honscheid}, {James},
  {Karliner}, {Kasen}, {Kent}, {Kuropatkin}, {Kuehn}, {Lahav}, {Li}, {Lima},
  {Maia}, {Margutti}, {Martini}, {Matheson}, {McMahon}, {Metzger}, {Miller},
  {Miquel}, {Mohr}, {Nichol}, {Nord}, {Ogando}, {Peoples}, {Plazas},
  {Quataert}, {Romer}, {Roodman}, {Rykoff}, {Sanchez}, {Scarpine}, {Schindler},
  {Schubnell}, {Sevilla-Noarbe}, {Sheldon}, {Smith}, {Smith}, {Smith},
  {Stebbins}, {Sutton}, {Swanson}, {Tarle}, {Thaler}, {Thomas}, {Tucker},
  {Vikram}, {Wechsler}, {Weller}, \& {DES Collaboration}}]{DECamgw150914}
{Soares-Santos}, M., {et~al.} 2016, \apjl, 823, L33

\bibitem[{{Soares-Santos} {et~al.}(2017){Soares-Santos}, {Holz}, {Annis},
  {Chornock}, {Herner}, {Berger}, {Brout}, {Chen}, {Kessler}, {Sako}, {Allam},
  {Tucker}, {Butler}, {Palmese}, {Doctor}, {Diehl}, {Frieman}, {Yanny}, {Lin},
  {Scolnic}, {Cowperthwaite}, {Neilsen}, {Marriner}, {Kuropatkin}, {Hartley},
  {Paz-Chinch{\'o}n}, {Alexander}, {Balbinot}, {Blanchard}, {Brown}, {Carlin},
  {Conselice}, {Cook}, {Drlica-Wagner}, {Drout}, {Durret}, {Eftekhari}, {Farr},
  {Finley}, {Foley}, {Fong}, {Fryer}, {Garc{\'{\i}}a-Bellido}, {Gill},
  {Gruendl}, {Hanna}, {Kasen}, {Li}, {Lopes}, {Louren{\c c}o}, {Margutti},
  {Marshall}, {Matheson}, {Medina}, {Metzger}, {Mu{\~n}oz}, {Muir}, {Nicholl},
  {Quataert}, {Rest}, {Sauseda}, {Schlegel}, {Secco}, {Sobreira}, {Stebbins},
  {Villar}, {Vivas}, {Walker}, {Wester}, {Williams}, {Zenteno}, {Zhang},
  {Abbott}, {Abdalla}, {Banerji}, {Bechtol}, {Benoit-L{\'e}vy}, {Bertin},
  {Brooks}, {Buckley-Geer}, {Burke}, {Carnero Rosell}, {Carrasco Kind},
  {Carretero}, {Castander}, {Crocce}, {Cunha}, {D{\rsquo}Andrea}, {da Costa},
  {Davis}, {Desai}, {Dietrich}, {Doel}, {Eifler}, {Fernandez}, {Flaugher},
  {Fosalba}, {Gaztanaga}, {Gerdes}, {Giannantonio}, {Goldstein}, {Gruen},
  {Gschwend}, {Gutierrez}, {Honscheid}, {Jain}, {James}, {Jeltema}, {Johnson},
  {Johnson}, {Kent}, {Krause}, {Kron}, {Kuehn}, {Kuhlmann}, {Lahav}, {Lima},
  {Maia}, {March}, {McMahon}, {Menanteau}, {Miquel}, {Mohr}, {Nichol}, {Nord},
  {Ogando}, {Petravick}, {Plazas}, {Romer}, {Roodman}, {Rykoff}, {Sanchez},
  {Scarpine}, {Schubnell}, {Sevilla-Noarbe}, {Smith}, {Smith}, {Suchyta},
  {Swanson}, {Tarle}, {Thomas}, {Thomas}, {Troxel}, {Vikram}, {Wechsler},
  {Weller}, {Dark Energy Survey}, \& {Dark Energy Camera GW-EM
  Collaboration}}]{GW170817DECampaper}
---. 2017, \apjl, 848, L16

\bibitem[{{Sorce} {et~al.}(2014){Sorce}, {Tully}, {Courtois}, {Jarrett},
  {Neill}, \& {Shaya}}]{sor14}
{Sorce}, J.~G., {Tully}, R.~B., {Courtois}, H.~M., {Jarrett}, T.~H., {Neill},
  J.~D., \& {Shaya}, E.~J. 2014, \mnras, 444, 527

\bibitem[{{Springob} {et~al.}(2014){Springob}, {Magoulas}, {Colless}, {Mould},
  {Erdo{\u g}du}, {Jones}, {Lucey}, {Campbell}, \& {Fluke}}]{spr14}
{Springob}, C.~M., {et~al.} 2014, \mnras, 445, 2677

\bibitem[{{Tanaka} \& {Hotokezaka}(2013)}]{tanaka13}
{Tanaka}, M., \& {Hotokezaka}, K. 2013, \apj, 775, 113

\bibitem[{{Tanaka} {et~al.}(2014){Tanaka}, {Hotokezaka}, {Kyutoku}, {Wanajo},
  {Kiuchi}, {Sekiguchi}, \& {Shibata}}]{tanaka14}
{Tanaka}, M., {Hotokezaka}, K., {Kyutoku}, K., {Wanajo}, S., {Kiuchi}, K.,
  {Sekiguchi}, Y., \& {Shibata}, M. 2014, \apj, 780, 31

\bibitem[{{Tanaka} {et~al.}(2017){Tanaka}, {Utsumi}, {Mazzali}, {Tominaga},
  {Yoshida}, {Sekiguchi}, {Morokuma}, {Motohara}, {Ohta}, {Kawabata}, {Abe},
  {Aoki}, {Asakura}, {Baar}, {Barway}, {Bond}, {Doi}, {Fujiyoshi}, {Furusawa},
  {Honda}, {Itoh}, {Kawabata}, {Kawai}, {Kim}, {Lee}, {Miyazaki}, {Morihana},
  {Nagashima}, {Nagayama}, {Nakaoka}, {Nakata}, {Ohsawa}, {Ohshima}, {Okita},
  {Saito}, {Sumi}, {Tajitsu}, {Takahashi}, {Takayama}, {Tamura}, {Tanaka},
  {Terai}, {Tristram}, {Yasuda}, \& {Zenko}}]{tanaka17gw170817}
{Tanaka}, M., {et~al.} 2017, \pasj, 69, 102

\bibitem[{{Taylor} \& {Weisberg}(1982)}]{tay82}
{Taylor}, J.~H., \& {Weisberg}, J.~M. 1982, \apj, 253, 908

\bibitem[{{The LIGO Scientific Collaboration} \& {the Virgo
  Collaboration}(2017{\natexlab{a}})}]{GW170817first}
{The LIGO Scientific Collaboration}, \& {the Virgo Collaboration}.
  2017{\natexlab{a}}, GRB Coordinates Network, Circular Service, No.~21505, \#1
  (2017), 1505

\bibitem[{{The LIGO Scientific Collaboration} \& {the Virgo
  Collaboration}(2017{\natexlab{b}})}]{GW170817detection}
---. 2017{\natexlab{b}}, GRB Coordinates Network, Circular Service, No.~21509,
  \#1 (2017), 1509

\bibitem[{{Theureau} {et~al.}(2007){Theureau}, {Hanski}, {Coudreau}, {Hallet},
  \& {Martin}}]{the07}
{Theureau}, G., {Hanski}, M.~O., {Coudreau}, N., {Hallet}, N., \& {Martin},
  J.-M. 2007, \aap, 465, 71

\bibitem[{{Tully}(1988)}]{tul88}
{Tully}, R.~B. 1988, {Nearby galaxies catalog}

\bibitem[{{Utsumi} {et~al.}(2017{\natexlab{a}}){Utsumi}, {Tominaga}, {Tanaka},
  {Morokuma}, {Yoshida}, {Asakura}, {Finet}, {Furusawa}, {Kawabata}, {Liu},
  {Matsubayashi}, {Moritani}, {Motohara}, {Nakata}, {Ohta}, {Terai}, {Uemura},
  {Yasuda}, \& {on behalf of the J-GEM collaboration}}]{uts17gw151226}
{Utsumi}, Y., {et~al.} 2017{\natexlab{a}}, ArXiv e-prints, arXiv:1710.00127

\bibitem[{{Utsumi} {et~al.}(2017{\natexlab{b}}){Utsumi}, {Tanaka}, {Tominaga},
  {Yoshida}, {Barway}, {Nagayama}, {Zenko}, {Aoki}, {Fujiyoshi}, {Furusawa},
  {Kawabata}, {Koshida}, {Lee}, {Morokuma}, {Motohara}, {Nakata}, {Ohsawa},
  {Ohta}, {Okita}, {Tajitsu}, {Tanaka}, {Terai}, {Yasuda}, {Abe}, {Asakura},
  {Bond}, {Miyazaki}, {Sumi}, {Tristram}, {Honda}, {Itoh}, {Itoh}, {Kawabata},
  {Morihana}, {Nagashima}, {Nakaoka}, {Ohshima}, {Takahashi}, {Takayama},
  {Aoki}, {Baar}, {Doi}, {Finet}, {Kanda}, {Kawai}, {Kim}, {Kuroda}, {Liu},
  {Matsubayashi}, {Murata}, {Nagai}, {Saito}, {Saito}, {Sako}, {Sekiguchi},
  {Tamura}, {Tanaka}, {Uemura}, \& {Yamaguchi}}]{utsumi17}
---. 2017{\natexlab{b}}, \pasj, 69, 101

\bibitem[{{Valenti} {et~al.}(2017){Valenti}, {David}, {Sand}, {Yang},
  {Cappellaro}, {Tartaglia}, {Corsi}, {Jha}, {Reichart}, {Haislip}, \&
  {Kouprianov}}]{valenti17}
{Valenti}, S., {et~al.} 2017, \apjl, 848, L24

\bibitem[{{Wanajo} {et~al.}(2014){Wanajo}, {Sekiguchi}, {Nishimura}, {Kiuchi},
  {Kyutoku}, \& {Shibata}}]{wanajo14}
{Wanajo}, S., {Sekiguchi}, Y., {Nishimura}, N., {Kiuchi}, K., {Kyutoku}, K., \&
  {Shibata}, M. 2014, \apjl, 789, L39

\bibitem[{{Willick} {et~al.}(1997){Willick}, {Courteau}, {Faber}, {Burstein},
  {Dekel}, \& {Strauss}}]{wil97}
{Willick}, J.~A., {Courteau}, S., {Faber}, S.~M., {Burstein}, D., {Dekel}, A.,
  \& {Strauss}, M.~A. 1997, \apjs, 109, 333

\bibitem[{{Yamazaki} {et~al.}(2016){Yamazaki}, {Asano}, \& {Ohira}}]{yam16}
{Yamazaki}, R., {Asano}, K., \& {Ohira}, Y. 2016, Progress of Theoretical and
  Experimental Physics, 2016, 051E01

\bibitem[{{Yoshida} {et~al.}(2017){Yoshida}, {Utsumi}, {Tominaga}, {Morokuma},
  {Tanaka}, {Asakura}, {Matsubayashi}, {Ohta}, {Abe}, {Chimasu}, {Furusawa},
  {Itoh}, {Itoh}, {Kanda}, {Kawabata}, {Kawabata}, {Koshida}, {Koshimoto},
  {Kuroda}, {Moritani}, {Motohara}, {Murata}, {Nagayama}, {Nakaoka}, {Nakata},
  {Nishioka}, {Saito}, {Terai}, {Tristram}, {Yanagisawa}, {Yasuda}, {Doi},
  {Fujisawa}, {Kawachi}, {Kawai}, {Tamura}, {Uemura}, \&
  {Yatsu}}]{JGEMgw151226}
{Yoshida}, M., {et~al.} 2017, \pasj, 69, 9

\end{thebibliography}

\end{document}